%% file: 4c7426-kmb-dec.tex
\documentclass[useAMS,usegraphicx,usenatbib]{mn2e} 

\usepackage{ulem}
\usepackage{color}
\input{rgb}

\newcommand{\gtapprox}{\raisebox{-0.5ex}{$\,\stackrel{>}{\scriptstyle
\sim}\,$}}
\newcommand{\ltapprox}{\raisebox{-0.5ex}{$\,\stackrel{<}{\scriptstyle
\sim}\,$}}

\usepackage{graphicx}
\usepackage{times} 
\usepackage{amssymb}
\usepackage{amsmath}
\usepackage{lscape}
\usepackage{url}
\newif\ifAMStwofonts
\AMStwofontstrue

\def\apj{ApJ}
\def\mnras{MNRAS}

\def\araa{ARA\&A}                
\def\aap{A\&A}                   
\def\aj{AJ}                      
\def\apjl{ApJ}                   

\def\chandra{{\it Chandra}}

\def\xmm{{\it XMM-Newton}}

\def\frii{\hbox{\rm FR\thinspace II}}

\def\ks{\hbox{$\rm\thinspace ks$}}
\def\xks{\hbox{$\rm ks$}}

\def\hz{\hbox{$\rm\thinspace Hz$}}

\def\ghz{\hbox{$\rm\thinspace GHz$}}

\def\kpc{\hbox{$\rm\thinspace kpc$}}
\def\mpc{\hbox{$\rm\thinspace Mpc$}}

\def\as{\hbox{$\rm\thinspace arcsec$}}
\def\am{\hbox{$\rm\thinspace arcmin$}}

\def\pcmsq{\hbox{$\rm\thinspace cm^{-2}$}}
\def\kmpspmpc{\hbox{$\rm\thinspace km~s^{-1}~Mpc^{-1}$}}
\def\assq{\hbox{$\mathrm{~arcsec^{2}}$}}

\def\kev{\hbox{$\rm\thinspace keV$}}

\def\mjypb{\hbox{$\rm\thinspace mJy/beam$}}
\def\jy{\hbox{$\rm\thinspace Jy$}}

\def\cts{\hbox{$\rm\thinspace counts$}}

\def\photpkevpcmsqps{\hbox{$\rm\thinspace photon~keV^{-1}~cm^{-2}~s^{-1}$}}

\def\ergpcmsqps{\hbox{$\rm\thinspace erg~cm^{-2}~s^{-1}$}}

\def\ergpcmsqpspassq{\hbox{$\mathrm{~erg~cm^{-2}~s^{-1}~arcsec^{-2}}$}}
\def\ergps{\hbox{$\rm\thinspace erg~s^{-1}$}}

\begin{document}

\title[Two different shocks within the hotspot of 4C74.26] 
{Two types of shock in the hotspot of the giant quasar 4C74.26: 
a high-resolution comparison from Chandra, Gemini \& MERLIN}  
\author[M. C. Erlund et al.]
{\parbox[]{6.in} {M.C.\,Erlund,$^{1}$ A.C.\,Fabian,$^{1}$ Katherine~M.\,Blundell,$^{2}$ 
C.S.\,Crawford$^1$\\ and P.\,Hirst$^3$ }\\\\
  \footnotesize
  $^{1}$Institute of Astronomy, Madingley Road, Cambridge CB3 0HA\\
  $^{2}$University of Oxford, Astrophysics, Keble Road, Oxford OX1 3RH\\
  $^{3}$Gemini Observatory Northern Operations Center, 670 N. A'ohoku
  Place, Hilo, Hawaii, 96720, USA}
\maketitle

\begin{abstract}
  New Chandra observations have resolved the structure of the X-ray
  luminous southern hotspot in the giant radio quasar 4C74.26 into two
  distinct features.  The nearer one to the nucleus is an
  extremely luminous peak, extended some 5\,kpc perpendicular to the
  orientation of the jet; 19\,kpc projected further away from the
  central nucleus than this is an arc having similar symmetry.  This
  X-ray arc is co-spatial with near-IR and optical emission imaged
  with Gemini, and radio emission imaged with MERLIN.  We explore how
  this double feature corresponds to two shocks having very different
  characteristics in spectral energy distribution.  We present the
  case that these observations are explained by the luminous X-ray
  peak being synchrotron emission with a flux density of $\sim
  7$\,nano-Jy.  There is no steep spectrum radio, optical or near-IR
  emission directly associated with this shock.  Beyond this point in
  the jet's flow, and following adiabatic losses, at least 19\kpc\
  further downstream where the flow impinges on the inter-galactic
  medium, the arc structure seen in sharp focus at radio wavelengths
  appears to be approximately mimicked at near-IR, optical and X-ray
  wavelengths.  The radio emission is most naturally explained as
  synchrotron emission but it would be unnatural to explain the X-rays
  as arising from the same emission mechanism since they show a
  non-monotonic spectrum.  They are, however, explicable as inverse
  Compton scattering of photons in the Cosmic Microwave Background,
  though this requires that the magnetic field strength in this
  slender arc/shock region responsible for the associated synchrotron
  radio emission is just a few per cent of the minimum energy value.
  The angular separation of the double shock structure (itself
  \gtapprox\ 19\,kpc or 10\,arcsec in size) from the active nucleus
  which fuels them of $\sim 550\,$kpc could present a challenge for
  connecting ``unidentified'' hard X-ray or Fermi sources with their
  origins.

\end{abstract}


\section{Introduction}

The proximity ($z = 0.104$) and size ($> 1.1$\,Mpc) of the giant radio
quasar 4C74.26 in principle mean that we should get a zoomed-in view
of structures within it, such as hotspots.  There are suggestions
  in the literature that hotspot size scales with the size of the
  radio source \citep[e.g.\ ][]{Machalski2008} meaning that giant
  radio sources can be profitable to study in order to investigate the
  hotspot phenomena, although this scaling dependence seems fairly
  weak \citep[e.g.\ ][]{BlundellRawlingsWillott}.   Hotspots in
classical double \frii\ \citep{fr} radio sources are believed to
represent the sites of particle acceleration. This is important as it
determines the spectrum (including the high- and low-energy turnovers)
of the particles that eventually find their way into the lobes of
these sources.  This in turn determines how much energy they inject
into their environments.

Multi-wavelength observations of \frii\ radio galaxies have shown that
low (radio) luminosity \frii\ sources typically emit optical and X-ray
synchrotron emission from their hotspots, whereas high luminosity
\frii\ sources do not \citep{meisenheimer89, hardcastle04}.  Like many
giant radio sources, 4C\,74.26 has a low luminosity \frii\ structure.
Contrary to expectation, our initial study of 4C\,74.26, a radio-loud
quasar at $z=0.104$, illustrated that the brightest X-ray hotspot
component does not necessarily trace the location of the brightest
radio feature (\citealt{erlund07}; see also \citealt{hardcastle07,
  goodger08, evans08}).  Our initial study of the southern hotspot of
4C\,74.26 showed it to have the largest known offset between its radio
and X-ray peak.  We investigated the causes of this offset in the
context of different emission processes and models. The initial X-ray
(archival \xmm\ and \chandra\ data, both 5\am\ off-axis and the latter
with gratings in place) and radio (VLA and MERLIN) observations were
well-modelled with either the dentist's drill model \citep{scheuer82},
inverse Compton upscattering of the cosmic microwave background in a
decelerating jet \citep{georganopouloskazanas04} or a simple
spine-sheath model \citep{chiaberge00}.  The fine spatial resolution
of the new data allows us to test these models and better constrain
the processes taking place at the hotspot.  Our analysis of the
archival X-ray observations of 4C\,74.26 presented in
\citet{erlund07}, showed that its southern hotspot is more X-ray
luminous than the western hotspot of Pictor\,A \citep{wilson01} and
hotspot\,D in Cygnus\,A \citep{wilson00}. It is also at a far greater
distance from its central nucleus than the hotspots in either of these
two sources, making 4C\,74.26 one of the largest sources in the Local
Universe with a projected linear size between its north and south
hotspot complexes of $1.1$\mpc\ \citep{pearson92}.

We present a multi-wavelength follow-up study of the multi-component
southern hotspot of 4C\,74.26 using new deep pointed observations from
\chandra\ and GEMINI (in $K$ band with NIRI
and $g'$ band with GMOS) together with some MERLIN data previously
published in \citet{erlund07}.

Throughout this paper, all errors are quoted at $1\sigma$ unless
otherwise stated and the cosmology is $H_{\rm 0} = 71$\kmpspmpc,
$\Omega_{0}=1$ and $\Lambda_{0} = 0.73$.  One arcsecond represents
$1.887$\kpc\ on the plane of the sky at the redshift of 4C\,74.26 and
the Galactic absorption along the line-of-sight towards this object is
$1.19 \times 10^{21}$\pcmsq\ \citep{dickeylockman90}.

\section{Data reduction}
\begin{table} 
\center
\begin{tabular}{lll}
\hline
Date       & ObsID & ksec   \\
\hline
2008-01-06 & $9231$ & $23.8$ \\
2008-01-23 & $9809$ & $12.0$ \\
2008-01-25 & $9800$ & $12.9$ \\
\hline
\end{tabular}
\caption[Details of the new \chandra\ observations of 4C\,74.26]
  {Details of our \chandra\ observations: (1) date of the observation, 
  (2) observation identification (ObsID) number 
  and (3) duration of observations; none of the observations suffered from flaring.}
\label{table:newchanobs} 
\end{table}

\subsection{New Chandra data}
\begin{figure*}
\center
\rotatebox{0}{
\resizebox{!}{22cm}
{\includegraphics{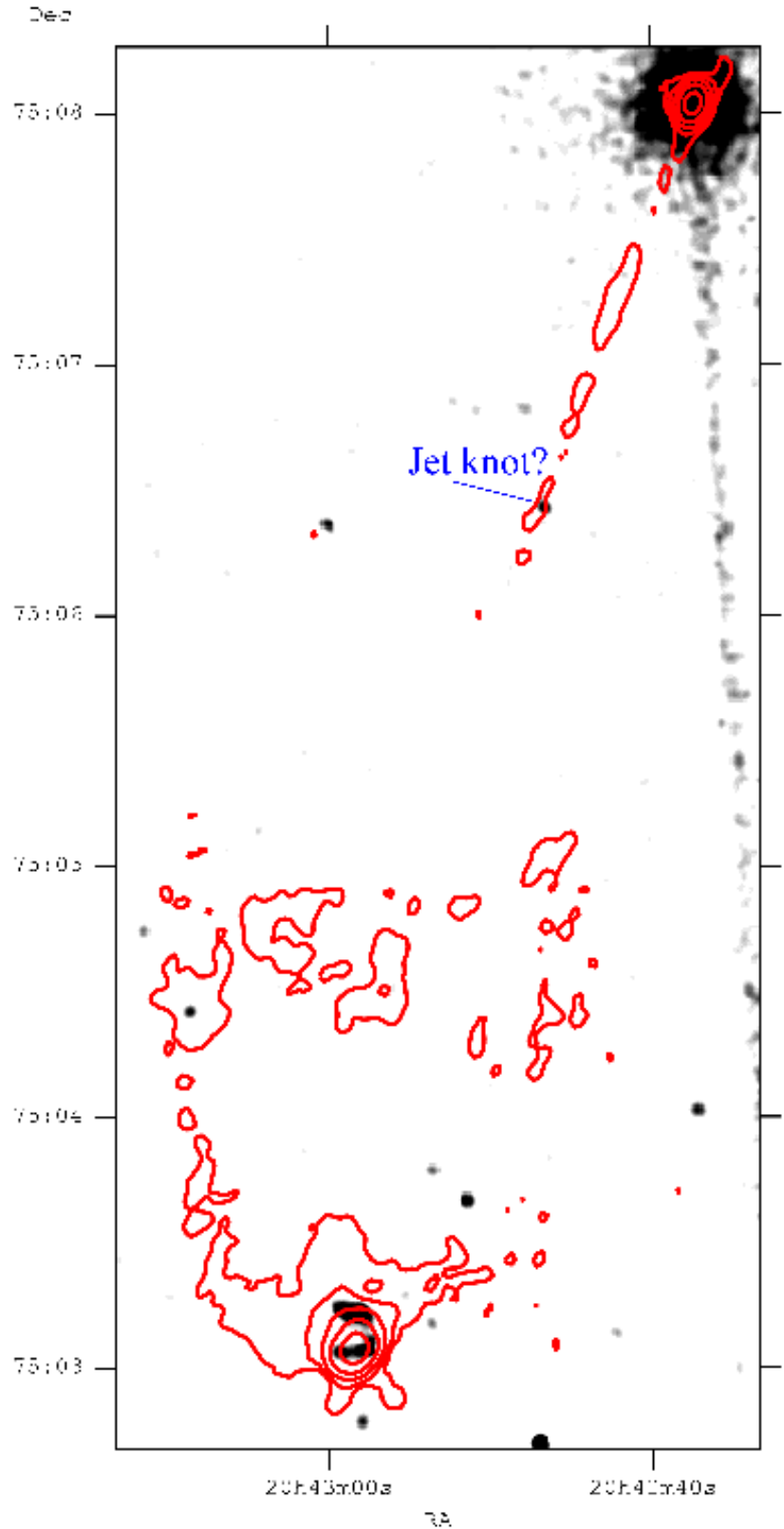}}
}
\caption[New \chandra\ data of the southern side of 4C\,74.26 with VLA
B-array contours overlaid]{\chandra\ data of the southern hotspot of
  4C\,74.26 in the $0.3-5$\kev\ band smoothed by a Gaussian kernel of
  $1.0$\as. The red contours are the 1.4\,GHz data from the VLA in
  B-array ($0.31$, $1.3$, $5.6$, $24$ and $100$\mjypb).  The label
  refers the X-ray source co-incident with the radio jet which is a
  candidate jet knot.  The grey streak is the \chandra\ readout
  streak.}
\label{fig:newfull}
\end{figure*}

\begin{figure*}
\center
\rotatebox{0}{
\resizebox{!}{8cm}
{\includegraphics{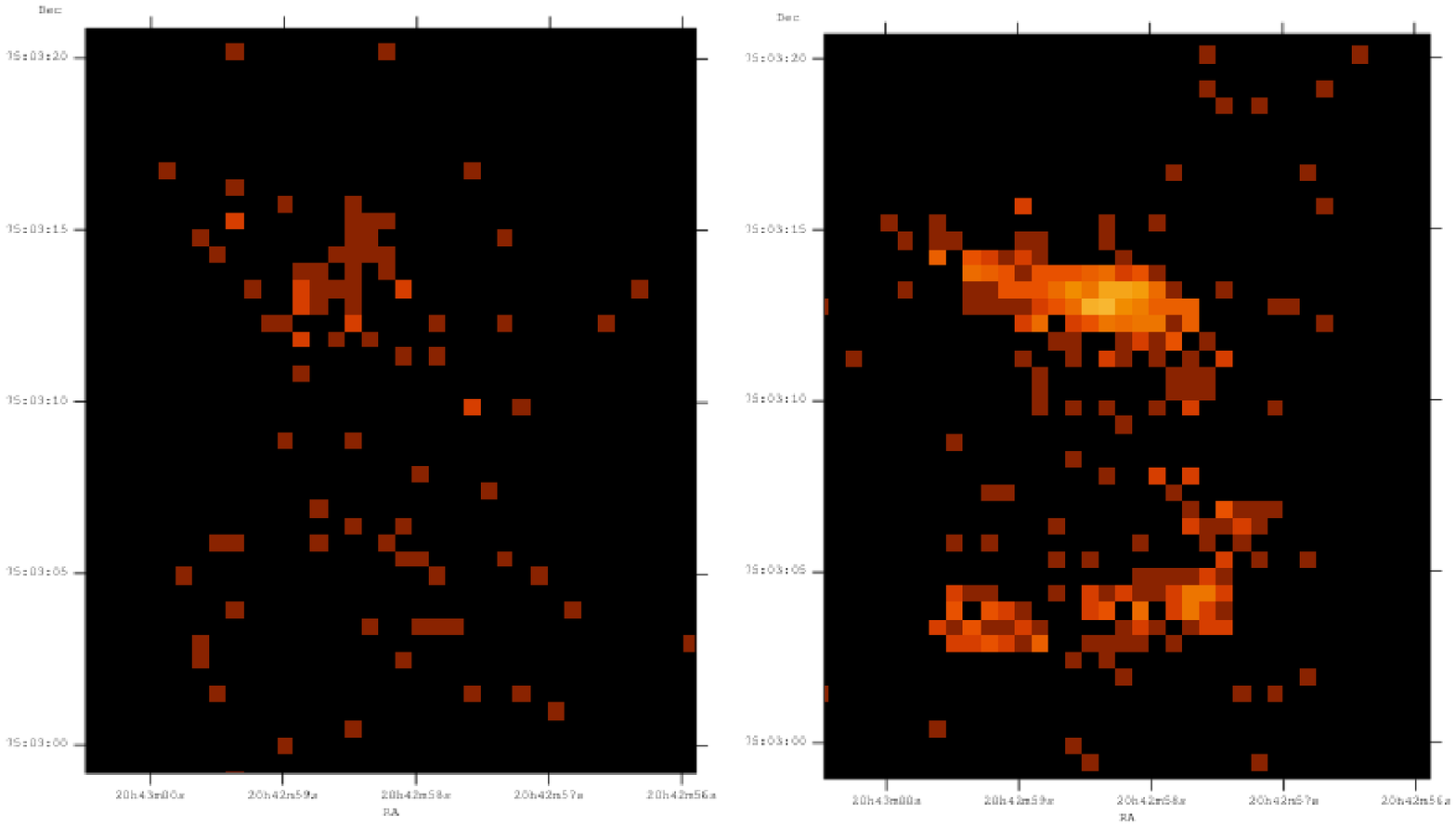}}
}
\caption[Comparison between the old and new \chandra\ data]{\chandra\
  data of the southern hotspot of 4C\,74.26 in the $0.3-5$\kev\
  band. {\it Left-hand side:} shows the original \chandra\ data 5\am\
  off-axis with the gratings in place.  {\it Right-hand side:} the
  new \chandra\ data.  The diffuse emission forms an arc of X-ray
  emission.  The original peak is see to be elongated.}
\label{fig:beforeafter}
\end{figure*}

The new \chandra\ data (observations listed in Table
\ref{table:newchanobs}) were processed using the {\sc
  acis\_process\_events} part of the {\sc ciao} data processing
software package ({\sc ciao}$\rm v.4.0.1$ and {\sc caldb}$\rm
v.3.4.2$).  Pixel randomisation was turned off for all observations
using the {\sc acis\_process\_events} tool. Then the Sub-pixel
Resolution Algorithm (\citealt{subpix} and \citealt{subpix2}) was used
to make use of photons that arrive near the edges and corners of the
pixels, as their arrival point can be determined with sub-pixel
resolution. All observations were free from flares giving a total of
$48.7$\ks\ of good quality data.  The data from ObsIDs $9800$ and
$9809$ were reprojected onto ObsID $9231$ (see Table
\ref{table:newchanobs}) as it is the longest observation.  The
observations were added together to produce a summed image.  The X-ray
images presented here use the summed data.  The northern hotspot lies
off the chip in all observations.  Fig. \ref{fig:newfull} shows the
new \chandra\ data overlaid with 1.4\,GHz data from the VLA in its
B-configuration.  The southern hotspot complex is clearly detected by
\chandra\ and in addition there is a source coincident with the jet
(marked `Jet knot?' in Fig.  \ref{fig:newfull}) that we consider to be
a candidate jet knot.  The core of 4C\,74.26 lies in the corner of the
S3 chip and suffers from significant pile-up.  The hotspot complex is
located $1.7$\am\ off-axis (this does not significantly affect the
spatial resolution of these observations) and this means that the path
of the southern jet is covered by the S3 chip.  The $90$\% enclosed
energy radius\footnote{Calculated using
  \url{http://cxc.harvard.edu/cgi-bin/build_viewer.cgi?psf}} at
1.5\kev\ at this distance off-axis is $\ltapprox 1.1$\as\ (on-axis it
is $0.94$\as).

Fig. \ref{fig:beforeafter} shows the difference in counts between the
original and new \chandra\ data.  The peak is now seen to be elongated
in the north-east---south-west direction and the diffuse emission is
clearly seen to trace an arc.  The marked difference between the two
data sets is due to the fact that the original data were taken with
the HETG in place, with the hotspot lying $5$\am\ off-axis; the new
data are from a $50$-\xks\ pointed observation of the southern hotspot
complex.  The morphology of the X-ray peak alone excludes the
possibility of it being a background AGN.  Despite the extended nature
of the components, {\sc ds9} was used to find the centroid of the peak
(RA 20h42m58.4s, Dec +75d03m12.8s), the southern arc (RA 20h42m58.2s,
Dec +75d03m04.5s) and the jet-knot candidate (RA 20h42m46.6s, Dec
+75d06m25.8s).  There are $286$\cts\ ($5$ in the background) in the
peak measured in an ellipse with a $6$\as\ semi-major axis and a
$3$\as\ semi-minor axis.  Another $132$\cts\ ($5$ in the background)
lie in the southern arc in a region the same size and shape as the
region used to calculate the number of counts in the peak but with a
different position and orientation.  A further $30$\cts\ ($1.6$\cts\
of which are background) are detected in the jet knot candidate in a
2\as-circular region.  All counts are measured in the $0.3-5$\kev\
band and summed over all three observations. These counts are
consistent with the original \chandra\ observation.  The jet is not
detected down to a $3\sigma$ upper limit of $9.5 \times
10^{-18}$\ergpcmsqpspassq (assuming a photon index of $\Gamma =
  1.7$), with the exception of the potential jet knot which lies
within the \xmm\ scattering spike (see \citealt{erlund07}).

The results of the spectral fits for the peak and arc regions as well
as the hotspot complex (to enable comparison with the \xmm\ spectral
fits) are shown in Table \ref{table:newchanspec}.  The new \chandra\
spectral data (spectra were extracted separately for each observation)
were fitted simultaneously for each region.  The \chandra\ spectra of
the hotspot complex region are consistent with the \xmm\
spectrum when fitted over the same spectral range. 

\subsection{Infrared and optical data}

Gemini North observed the southern hotspot complex of 4C\,74.26 (as
part of program GN-2008A-Q-102) with NIRI (on 2008 May 7 and 9) and
GMOS (on 2008 June 6) for a total of $7.02$\ks\ in the $K$ band and
$5.4$\ks\ in the $g'$ band respectively.  We used standard observing
sequences and performed the data reduction, including computing a deep
sky frame from the offset science frames using PyRAF and the Gemini
Data Reduction Package.  We then registered the GMOS data to the SDSS
catalogue and the NIRI data to the 2MASS catalogue in order to
astrometrically and flux calibrate the images.  The registration of
the SDSS frame [to which the GMOS data were aligned] with respect to
the radio frame has been established to be better than 0.2\,arcsec
\citep{Souchay2008}.  The X-ray image was aligned with respect to the
corrected Gemini images so we are confident that the overall alignment
of our images is good to a few tenths of an arcsecond and thus that
the offset plotted in Figure\,\ref{fig:nprofile} is real.  

\begin{figure*}
  \center \rotatebox{0}{ \resizebox{!}{9cm}
    {\includegraphics{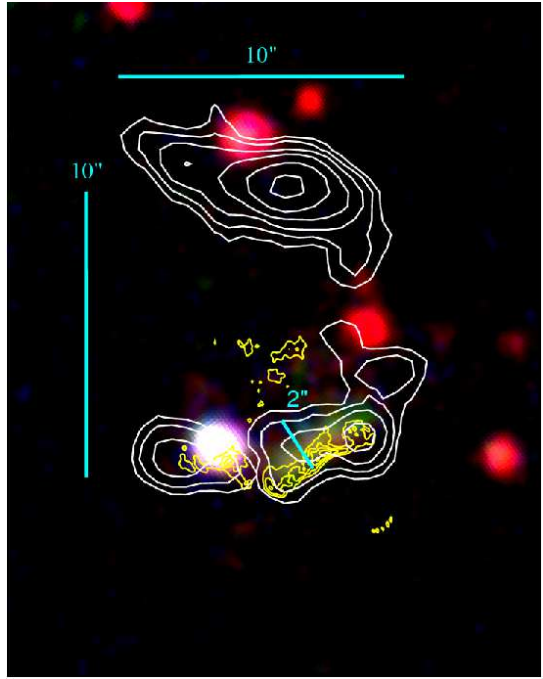}}
  }
  \caption[Multi-wavelength image of the southern hotspot complex in
  4C\,74.26]{NIRI $K$ band data are in red, GMOS $g'$ band data in
    green and WHT R band data are in blue.  The white contours are the
    new \chandra\ data and the yellow contours are MERLIN data.  
      Angular separations are shown in arcsec.}
\label{fig:firstlookgn}
\end{figure*}

The data show that the southern arc feature of the hotspot complex is
detected as faint diffuse emission in both $K$ and $g'$ bands (see
Fig.  \ref{fig:firstlookgn}) and has a clear linear structure
particularly in the ``west wing'', with some very faint emission
extended north towards the X-ray peak.  This linear structure (in both
bands) traces the brightest edge of the MERLIN radio emission,
seemingly cupped within it.  There is a point-like source close ($\sim
0.5$\as) to the easternmost peak in the MERLIN structure in both $K$
and $g'$ band images (where the FWHM seeing was $0.6''$ and $0.9''$
respectively). It is identified as the source detected at $2.4\sigma$
with the Liverpool Telescope \citep{erlund07}.  The spectral energy
distribution (SED) and unresolved nature of this source indicates that
it is probably a star.  There is also a source $2.3$\as\ from the
north X-ray peak, that is also likely to be a background source.
There is no convincing detection of near-IR or optical emission
coincident with the north X-ray peak down to the limit of our
observations.

The WHT data were collected on the 2007 July 11 and 27 for a total of
$6.9$\ks\ in the $R$ band and $5.7$\ks\ in the $B$ band respectively.
The hotspot complex is not detected in the $B$ band (where the FWHM
seeing was $1.5''$). The southern arc feature is detected in the $R$
band (where the FWHM seeing was $0.9''$), again with diffuse emission
matching that seen in the $K$ and $g'$ band images.  This $R$ band
emission also lies along the MERLIN emission appearing to be cupped
within it (see Fig. \ref{fig:firstlookgn}).  The source detected close
to the easternmost MERLIN peak in the $K$ and $g'$ band is also
detected in the $R$ band WHT data.  Again no source is detected
coincident with the north X-ray peak. The source offset by $\sim
2.3$\as\ detected in the Gemini observations is also detected with
WHT.

\section{Results}

\subsection{The hardness ratio of the X-ray hotspot components}
\label{sec:hr}
\begin{figure*}
\center
\rotatebox{0}{
\resizebox{!}{9cm}
{\includegraphics{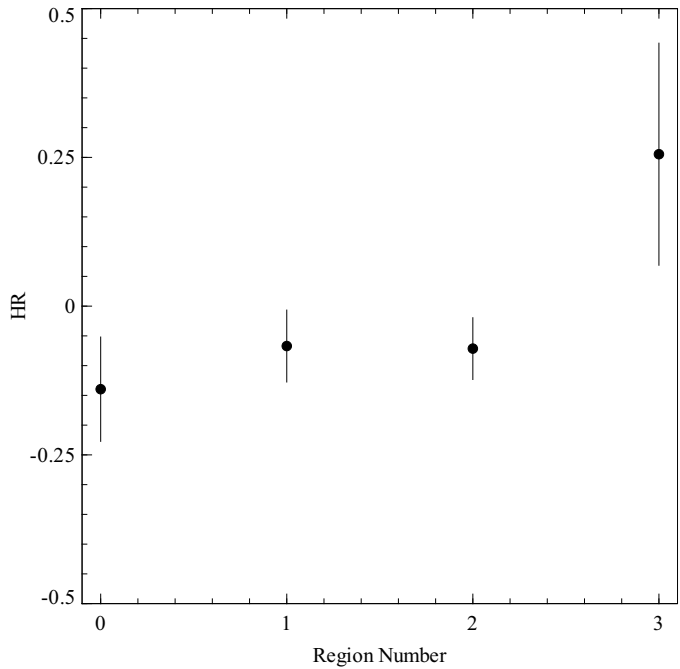}}
}
\caption[X-ray hardness ratios for the X-ray hotspot components and
jet knot candidate]{The hardness ratio of the arc (region number 0),
  the peak (region number 1), the hotspot complex (region number 2)
  and the jet knot candidate (region number 3).  The soft band is
  $0.5-1.5$\kev\ and the hard band is $1.5-7$\kev.}
\label{fig:hrnewchan}
\end{figure*}

\begin{figure*}
\center
\rotatebox{0}{
\resizebox{!}{9cm}
{\includegraphics{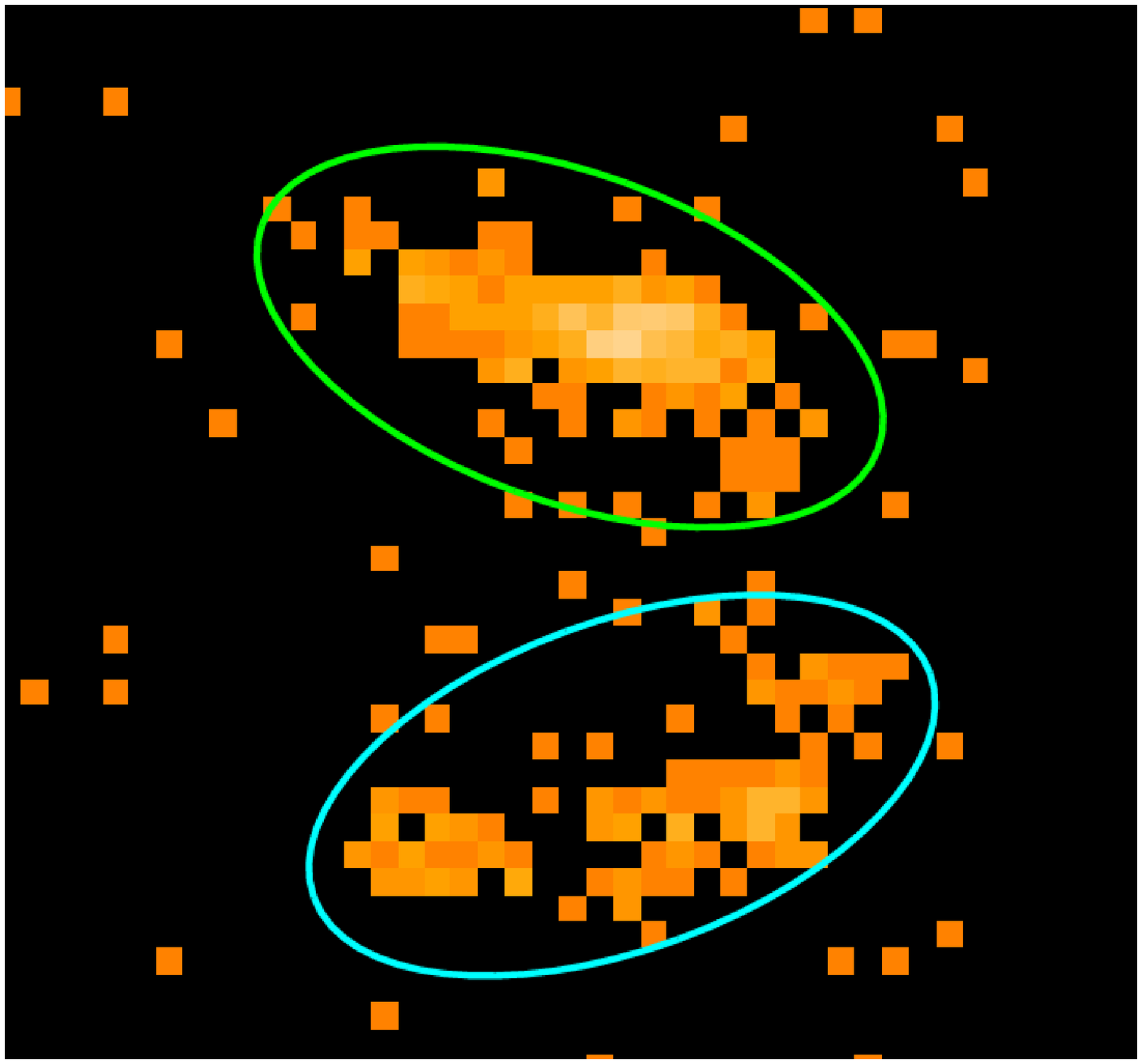}}
}
\caption[\chandra\ image illustrating the ellipses from which the
counts were measured.]{The ellipses used to measure the counts in the
  hard and soft bands of the Chandra data; the major axes are
  $12^{\prime\prime}$ and the minor axes are $6^{\prime\prime}$ in
  extent.  }
\label{fig:ellipses}
\end{figure*}

In order to better compare the two X-ray components of the southern
hotspot complex, their hardness ratios are measured.\footnote{The
  hardness ratio is defined as $HR = \frac{H-S}{H+S}$ where $H$ is the
  number of counts in the hard band and $S$ is the number of counts in
  the soft band.}  The number of X-ray counts in the hotspot
components, whole hotspot region and jet knot candidate were
calculated in the soft ($0.5-1.5$\kev) and hard ($1.5-7$\kev) band
respectively using {\sc CIAO} tool {\sc dmstat}. These energy bands
were chosen so that the number of counts in each band would be roughly
equal in ellipses with a major axis of $12^{\prime\prime}$ and a
  minor axis of $6^{\prime\prime}$ located and oriented as illustrated
  in Figure\,\ref{fig:ellipses}. The background was measured in a
  source-free region from a circle with radius $41.7^{\prime\prime}$.
  These values were treated as 'expected values' and generated arrays
  poisson distributed about each expected value.  Background
  subtracted counts were then calculated to obtain a hardness ratio
  with associated error bars.   The resulting hardness ratios for
the peak, arc, hotspot complex and jet knot candidate are shown in
Fig.  \ref{fig:hrnewchan}. There is no significant difference between
the hardness ratio of the peak and arc X-ray features.  The jet knot
candidate is however harder than the hotspot components.

\begin{figure*}
\center
\rotatebox{0}{
\resizebox{!}{9cm}
{\includegraphics{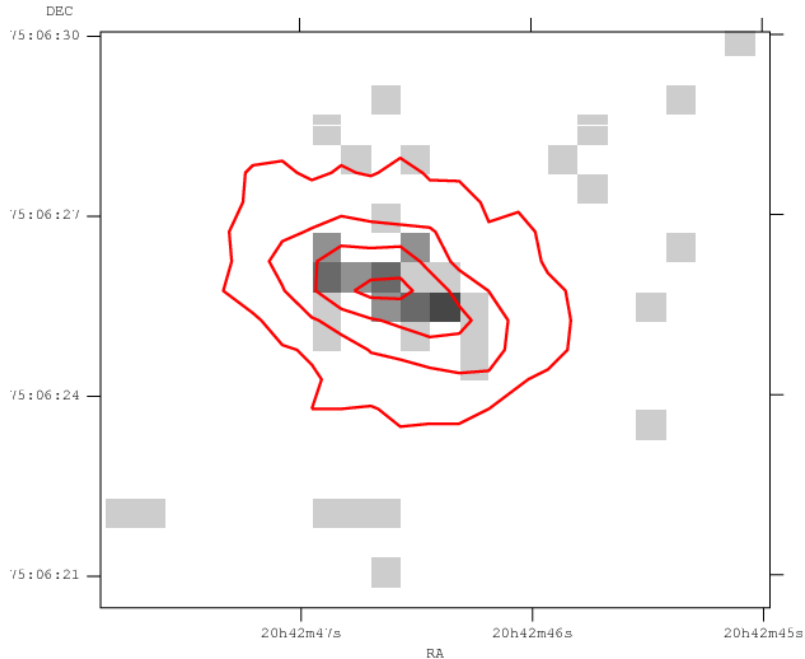}}
}
\caption[\chandra\ image of the jet knot candidate]{Grey scale image
  of the jet knot candidate in the $0.5-7$\kev\ band, with contours
  from a {\sc marx} ray trace simulation of a point source at the
  location of the knot candidate.  The jet knot candidate is
  consistent with an off axis point source (the contours are at 94\%,
  48\%, 17\% and 2\% of the peak pixel value in the {\sc marx}
  simulation).  }
\label{fig:marxknot}
\end{figure*}

\begin{figure*}
\center
\rotatebox{0}{
\resizebox{!}{9cm}
{\includegraphics{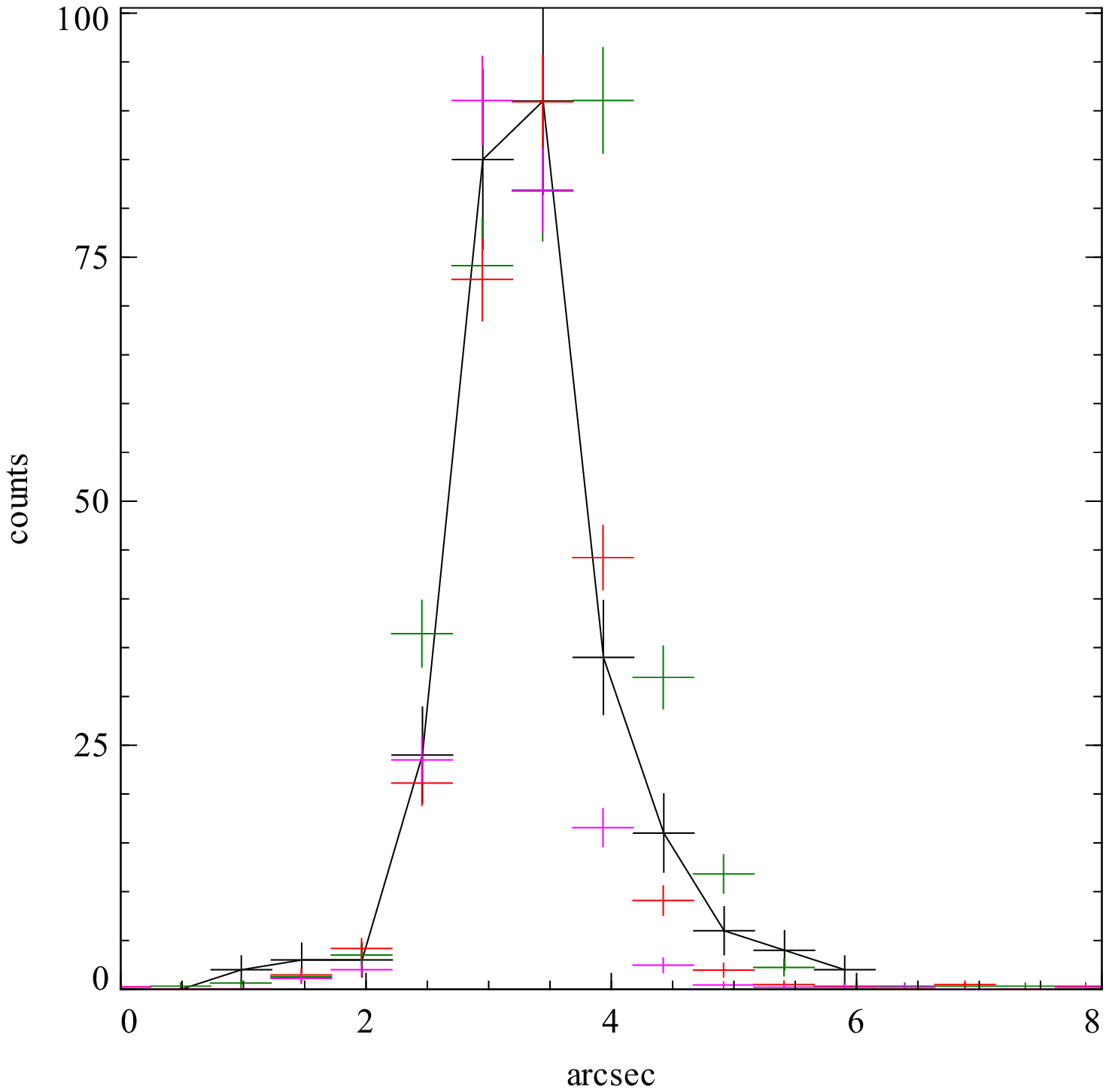}}
}
\caption[Is the northern X-ray hotspot component
resolved?]{North-to-south profile through the northern component of
  the hotspot complex is shown in black.  The data are in the
  $0.5-7$\kev\ band and the profile is constructed using a 10-\as\
  wide box.  The coloured points represent profiles through the {\sc
    marx} simulation of a line shaped source in the location of the
  northern peak of the hotspot complex with a similar inclination.
  The magenta points are for a profile at $90\deg$ to the line, the
  red points are $85\deg$ to the line and the green points are
  $80\deg$ to the line.  The northern hotspot is shown to be
  unresolved in the north--south direction.  Note the spectral model
  used is that of the \xmm\ data published in \citet{erlund07}.}
\label{fig:marxbar}
\end{figure*}

\subsection{Characterising the jet knot candidate}

Studying the jet-knot candidate in the new \chandra\ data in more
detail (Fig. \ref{fig:marxknot}), it appears to be extended
perpendicular to the jet axis.  However, it is $3.5$\am\ off-axis
($3.3$\am\ from the southern hotspot complex) and so a simulation
using {\sc marx}, a \chandra\ specific ray-tracing package, was
generated for a point source at this location in order to determine
whether its lateral extent is an effect of its off-axis position or
whether it has a similar bar-like morphology to the X-ray hotspot
components.  The results are shown in Fig.  \ref{fig:marxknot}, where
the contours are from the simulated point-spread function (PSF) and
trace the form of jet knot candidate extremely well. The source is
therefore consistent with an off-axis point source.  The spectral
model used in this simulation was generated by calculating the
relationship between hardness ratio (calculated above, see Fig.
\ref{fig:hrnewchan}) and photon index (taking into account Galactic
absorption).  This gives a photon index of $\Gamma = 1.03\pm 0.25$,
which is extremely flat (the errors were calculated by projecting the
$1\sigma$ errors on the hardness ratio onto the photon index
axis). The source is also detected in the Gemini GMOS $g'$ band image
as a faint source with a $g'$ magnitude of 24.88. Better data are
needed to discern whether or not this source is actually a jet knot,
or a background galaxy which hosts an obscured AGN.

\begin{figure*}
\center
\rotatebox{0}{
\resizebox{!}{9cm}
{\includegraphics{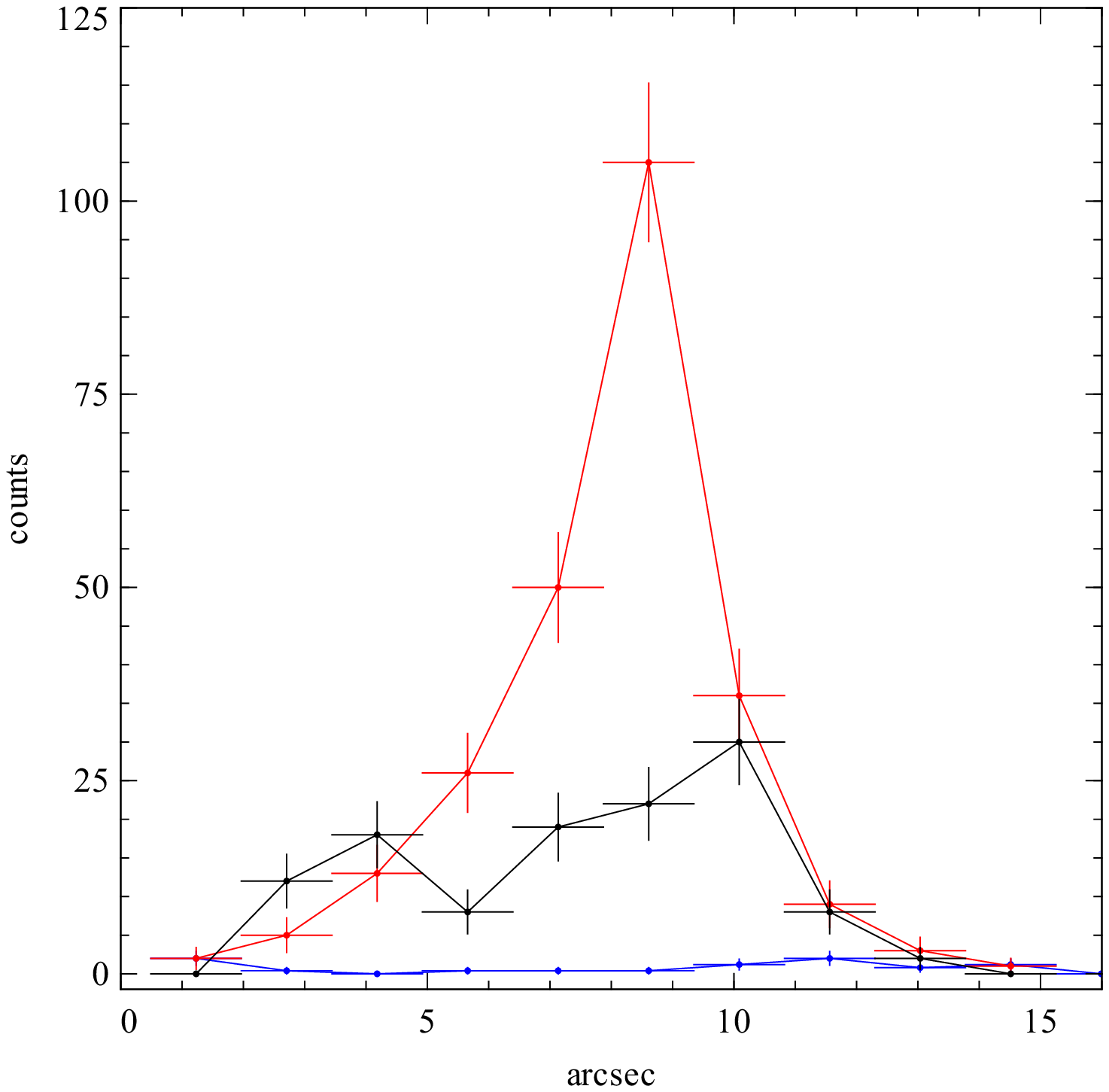}}
}
\caption[Transverse profiles through the north and south X-ray hotspot
components]{East-to-west profiles through the north and south X-ray
  components of the southern hotspot complex of 4C\,74.26.  The red
  line is the north X-ray peak and the black line is the south X-ray
  arc.  The blue line is the background profile scaled to take into
  account the narrow width of the box used for these lateral profiles.
  These profiles have been created using the unsmoothed $0.3-5$\kev\
  band \chandra\ data with counts per $4\times 1.5$\assq\ bin.  }
\label{fig:ewxprof}
\end{figure*}

\begin{figure*}
\center
\rotatebox{0}{
\resizebox{!}{9cm}
{\includegraphics{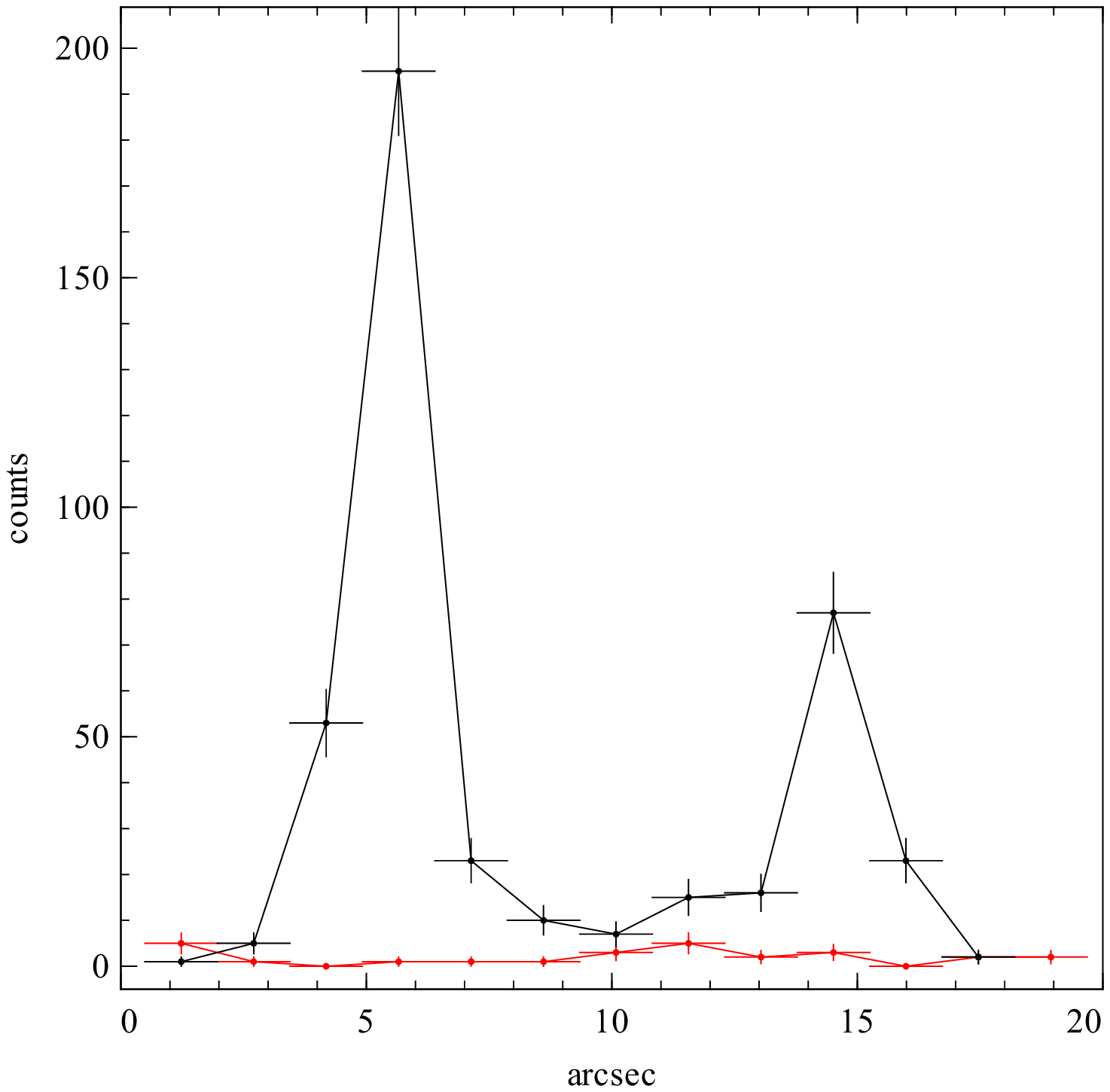}}
}
\caption[Profile from north to south through the new \chandra\
data]{Profile from north to south through southern hotspot complex of
  4C\,74.26 (black line) and through source-free region (red line).
  These profiles have been created using the unsmoothed $0.3-5$\kev\
  band \chandra\ data with counts per $10\times 1.5$\assq\ bin.  }
\label{fig:nsxprof}
\end{figure*}

\subsection{Resolving the X-ray hotspot peak}

Considering the X-ray hotspot complex more thoroughly, the
northernmost feature has been modelled (using {\sc marx}) as a source
in the form of a line at the same location and inclination on the
chip, as the observed X-ray peak in the longest observation ObsID 9231
(see Table \ref{table:newchanobs}).  Profiles were created across the
simulated line, perpendicular to it and at decreasing angles.  These
were re-normalised to compare them to the profile measured
perpendicular to the X-ray peak feature (see Fig.
\ref{fig:marxbar}).  Fig. \ref{fig:marxbar} shows that this feature
is not resolved in the (roughly) north--south direction.  However,
Fig. \ref{fig:ewxprof} shows that it is resolved from east to west,
which puts a constraint on the jet opening angle (see Section
\ref{sec:xpeak}).  Fig. \ref{fig:ewxprof} also shows that the
southern arc is resolved from east to west with a less peaked profile
than the northern X-ray component of the hotspot complex.  

Fig.  \ref{fig:nsxprof} shows the profile from north to south through
the X-ray peak and arc.  There is clearly no gentle gradation from the
initial X-ray peak to the southern arc, rather the two regions are
clearly separated with a small (but clearly detected) amount of
diffuse emission between them, as the background profile shows (in
red, in Fig. \ref{fig:nsxprof}). The background profile was measured
close to the hotspot complex using a similarly sized and orientated
box over a source-free region.

\begin{figure*}
\center
\rotatebox{0}{
\resizebox{!}{9cm}
{\includegraphics{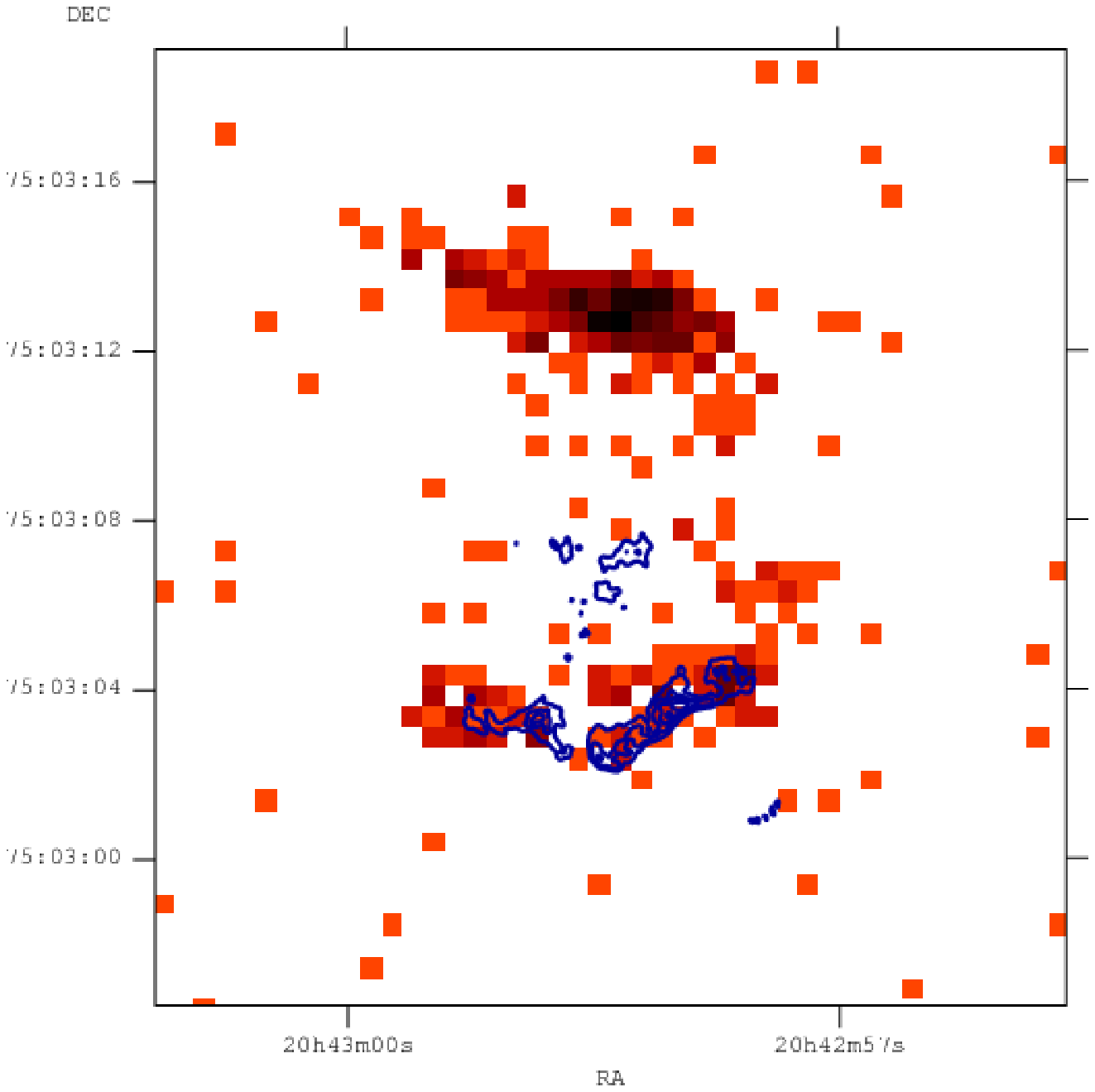}}
}
\caption[New \chandra\ data with MERLIN contours overlaid]{\chandra\
  data of the southern hotspot of 4C\,74.26 in the $0.3-5$\kev\ band,
  with MERLIN data overlaid (the contour levels are 0.5, 1, 1.5,
  2.0\mjypb)}
\label{fig:newchanmerl}
\end{figure*}

\begin{figure*}
\center
\rotatebox{0}{
\resizebox{!}{9cm}
{\includegraphics{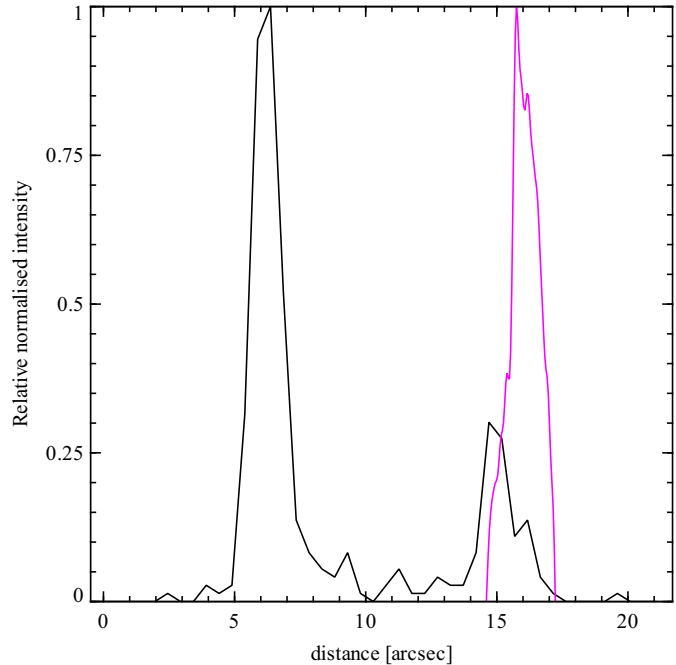}}
}
\caption[Profile through the new \chandra\ and MERLIN data of the
southern hotspot complex]{A profile through the new \chandra\ data of
  the southern hotspot complex from north to south (in the
  $0.3-5$\kev\ band, 5.4\as\ wide box) is shown in black.  The MERLIN
  data is in magenta. The error in the relative positioning of the
  profiles is less than $0.49$\as\ for \chandra\ and $0.05$\as\ for
  MERLIN}
\label{fig:nprofile}
\end{figure*}

Overlaying the new \chandra\ data with the MERLIN emission (see Fig.
\ref{fig:newchanmerl}) shows that the southern X-ray arc traces the
MERLIN emission, extending beyond it to the west by $\sim 3$\as\ and
to the east by $\sim 1$\as.  Fig. \ref{fig:nprofile} shows that the
southern X-ray peak is roughly coincident with the MERLIN peak.  The
southern arc is more powerful in X-rays than it is in radio emission:
$6.7\times 10^8$\jy\hz\ at $1$\kev\ as compared with $3.8\times
10^8$\jy\hz\ at $1.66$\ghz\ in the MERLIN data.

\section{Discussion}

\subsection{Emission processes in the hotspot components}

The north X-ray peak and south arc show very different broad band
spectra: the X-ray peak is the brightest X-ray region in the hotspot
complex but is not detected in any of the other observations at
different wavelengths; the arc is fainter in X-rays, but is detected
in the radio, infrared and optical bands.  Here, we comment on the
likely emission processes in these two components of the hotspot
complex.

\subsubsection{The northern X-ray peak}
\label{sec:xpeak}

\begin{figure*}
  \center \rotatebox{0}{ \resizebox{!}{9cm}
    {\includegraphics{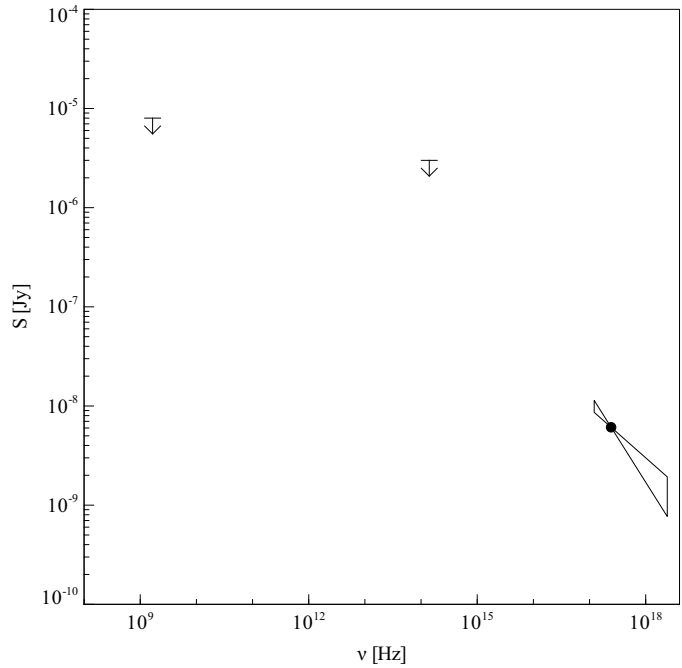}} }
  \caption[SED of northern X-ray hotspot component]{SED of the north
    X-ray peak in black, using the \chandra\ values for the spectral
    index and flux density.  The arrows represent the $3\sigma$ upper
    limit on any associated compact (1.66\ghz\ MERLIN) radio emission
    and $K$ band IR emission from the Gemini North NIRI data.}
  \label{fig:nsed}
\end{figure*}


In classical acceleration models such as those suggested by
\citet{meisenheimer89} (and references therein), the high-energy
synchrotron emission (X-ray / optical) reveals the location of the
Mach disk (inner shock).  The Mach disk is the most likely location
for efficient (first-order Fermi) acceleration. This requires a planar
shock which is a good description of the X-ray morphology of the
northern peak because it is resolved from east to west but not north
to south. The observed X-ray spectral index ($\alpha \sim 0.7$, from
\xmm\ and \chandra) is consistent with shock acceleration.
\citet{meisenheimer89} showed that low-luminosity FRII radio galaxies
generally tend to emit optical synchrotron emission at their hotspots
whereas more powerful FRII radio galaxies do not; this trend has
  been carefully investigated with more data by \citet{Brunetti2003}.
\citet{hardcastle04} confirmed that this trend continues up to X-ray
synchrotron wavelengths.  4C\,74.26 is a low-luminosity radio quasar
and so is expected to produce X-ray and optical synchrotron emission
in its hotspots.  An SED for the X-ray peak is consistent with it
being synchrotron emission within the constraints shown in
Fig.\,\ref{fig:nsed}.  The synchrotron interpretation for the
  northern component requires the presence of a break in the spectrum
  shown in Figure\,\ref{fig:nsed}.  Extrapolating from the X-ray
  emission to the upper limit at radio wavelengths would imply a
  spectral index no steeper than $\nu^{-0.4}$ so if the spectral index
  measured across the X-ray band ($\alpha \sim 0.7$) extends at all to
  shorter wavelengths than the X-rays, then a spectral break is
  required which may correspond to the elusive low-energy turnover in
  the synchrotron spectrum.  

Interestingly, classic acceleration models
(e.g. \citealt{meisenheimer89}) predict a double shock structure with
the inner shock at the Mach disk and an outer shock where the
intergalactic material is swept aside by the advancing jet head. The
structure they predict looks qualitatively similar to the morphology
of the X-ray hotspot complex in 4C\,74.26 except on much smaller
physical scales, than the south hotspot of 4C\,74.26 which at 19\,kpc
(projected) or 27\,kpc (deprojected) could comfortably accommodate
several M87 jets!

\citet{meisenheimer89} also point out that if the jet's magnetic field
strength increases, after the inner shock, towards where the flow
reverses direction due to interaction with the surrounding medium,
this can produce an offset between the high-energy synchrotron
emission at the Mach disk and low-energy synchrotron emission further
downstream.  However, as we will see in Section \ref{sec:constraints},
this model cannot explain the X-ray--radio peak offset in 4C\,74.26.

The lateral extent of the X-ray peak ($\sim 5-6$\as) places a
constraint on the opening angle of the jet given that the hotspot is
$5$\am\ from the core.  This means that for a source aligned on the
plane of the sky, the opening angle would be $\sim 1\deg$.  4C\,74.26
is a quasar, so assuming that it is aligned at $\le 45\deg$ to the
line of sight (the maximum for a quasar according to the unified model
of AGN), then the opening angle is $\ltapprox 0.5\deg$.  This implies
that the jet remains extremely well collimated over $\ge 800$\kpc\
from the core.

\subsubsection{The southern arc}
\label{sec:xarc}

\begin{figure*}
  \center \rotatebox{0}{ \resizebox{!}{9cm}
      {\includegraphics{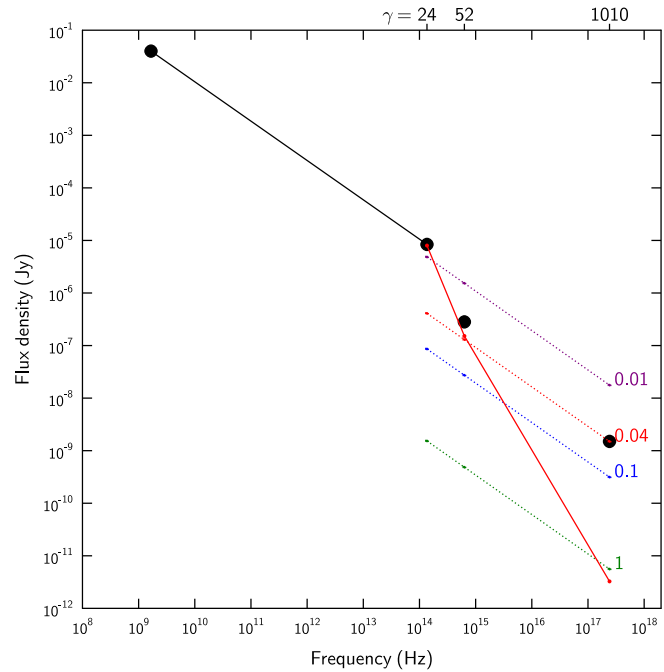}} }
    \caption[SED of southern X-ray hotspot component]{SED of the south
      X-ray arc.  The large solid black points mark the observed flux
      densities in the radio, IR, optical and X-ray bands for the western
      wing of the MERLIN emission only, because the eastern side of
      the arc is contaminated by a
      star.  
      The coloured sloping dotted lines represent the predicted ICCMB
      emission at the wavelengths shown if the magnetic field strength
      is lower than the minimum energy value by the factor shown to
      the right of each line.  The solid red line shows the difference
      between the observed flux density and the predicted ICCMB for
      the dotted red line (corresponding to 4\% of the minimum energy
      $B$-field strength) --- if our assumptions are correct then this
      line indicates the shape of the synchrotron spectrum at optical
      and higher energies.  The Lorentz factors of the particles
      responsible for the ICCMB at the three bands at which we have
      observations are plotted on the top $y$-axis.  }
  \label{fig:ssed}
\end{figure*}
The brightness structure of the southern arc in our radio, infra-red,
optical and X-ray observations broadly indicates that it is likely to
be produced from a dome-shaped shock.  In this regard, is is
remarkably similar to the dome-shaped shock (revealed in radio,
optical and infra-red wavelengths) of 3C\,445 by \citet{Prieto2002}
and more recently in X-rays by \citet{Perlman09}.  The SED of
the western half of the arc structure is plotted ins
Fig.\,\ref{fig:ssed} with the observed flux densities joined by the
black solid line.  Only the western half is considered in this
analysis because of the presence of a contaminating star on the
eastern side of the arc: the X-ray flux measurement has been scaled to
reflect the number of counts in the western wing of the MERLIN
emission as opposed to the whole X-ray arc region.  Even with this
careful definition of the region under consideration, it is
immediately apparent that the X-ray emission associated with the arc
is unlikely to be synchrotron because it gives a spectrum with a
non-monotonic gradient at these energies which would be
uncharacteristic, unless of course a multi-zone plasma is
  present.

  Therefore we investigated whether the emission could be explained as
  external Compton up-scattering from some source of seed photons.  In
  principle, these photons could arise from the synchrotron-emitting
  plasma responsible for the radio emission from this southern arc
  (synchrotron self-Compton, SSC), although the offset between
  the radio emission and the X-ray emission demonstrated in
  Figure\,\ref{fig:nprofile} makes this explanation unlikely in
  practice in this particular case.  Moreover, the flux density
    measured by MERLIN in the west wing of the southern arc
    corresponds, at a redshift of 0.104 to a radio luminosity of
    $1. \times 10^{41} {\mathrm{erg\,s^{-1}}}$ and hence an energy
    density of $8 \times 10^{-15} {\mathrm{erg\,cm^{-3}}}$
    considerably smaller than the energy density due to the CMB
    radiation field at this redshift which is $6 \times 10^{-13}
    {\mathrm{erg\,cm^{-3}}}$.  Thus, even if the synchrotron and X-ray
    emission were co-spatial, SSC would be small in comparision to
    ICCMB particularly when it is considered that SSC requires higher
    energy electrons than ICCMB and the spectrum is steep.
Therefore we consider inverse-Compton up-scattering of CMB photons
(hereafter ICCMB) from the same plasma population as is giving rise to
the synchrotron emission at radio wavelengths, albeit from particles
with lower Lorentz factors than are responsible for the radio
synchrotron observable with current radio telescopes.

The ratio of ICCMB emission and synchrotron emission from the same
population is governed by the magnetic field strength that is needed
for synchrotron emission to occur, given the number density
distribution of relativistic electrons that gives rise to both
emission mechanisms.  Tucker (1975) gives this relation (in
Equation\,4-53) and we plot in Fig.\,\ref{fig:ssed} the ICCMB emission
predicted on the basis of how much radio synchrotron emission is
observed for different fractions of the minimum energy magnetic field
strength of $1.7 \times 10^{-9}$\,T.  This value is calculated
assuming a $\gamma_{\rm min}$ of 10, a point to which we will return
later, and assuming a spectral dependence of $\nu^{-0.75}$ which is
the measured value of the spectral index between the radio and optical
data-points, and close to the inferred photon index of $\Gamma = 2.1
\pm 0.2$ (see Table\,\ref{table:newchanspec}).

Fig.\,\ref{fig:ssed} shows that if the population of particles
responsible for the synchrotron emission arises in a magnetic field
strength just $\sim 4$\,\% of the minimum energy value, then the ICCMB
emission predicted from the same population matches the flux density
we observe (red dotted line).  The figure also shows that the amount
of ICCMB emission predicted at the wavebands of our infra-red and
optical observations, when subtracted from the flux densities we
observe at these bands, gives a straight steep power-law
characteristic of aged synchrotron emission (shown as solid red line)
and a spectral break frequency at $\sim 10^{14}$\,Hz comparable with
those found in low power FRII radio sources by \citet{Mack2009}.

It is quite a step to invoke, in this slender region of the hotspot, a
$B$-field strength of $1.7 \times 10^{-9}$\,T when this is just $\sim
4$\,\% minimum energy value.  However, compression is manifestly
occurring in this region of the hotspot. It is possible, at least in
principle, that the timescales for compression of the magnetic field
strength are not well-matched with the timescales on which mixing, and
hence equipartition of the energies (which is close to the condition
for the minimum energy configuration), in the magnetic field and in
the particles has yet to take place.

We remark that it is only in the slender region delineated by the
MERLIN arc that we invoke a $B$-field so much lower than the minimum
energy value.  We emphasize that the relative resolution across this
hotspot in 4C74.26 surpasses that of many, if not all, other observed
hotspots published in the literature: this is partly because 4C74.26
is such a giant (at 1.1\,Mpc) and partly because it is so close to
Earth (at redshift 0.104).  It is possible that it has gone unnoticed
previously that very localised regions of hotspots (whose natures are
widely acknowledged to be complex) are out of equipartition simply
because the resolution has been insufficient (and/or the wavelength
coverage has been inadequate) to discern what is going on.



\subsubsection{Diffuse emission associated with the south hotspot}

The long baselines of the MERLIN interferometer filter out smooth,
extended emission that is detected by the much shorter baselines of
the VLA in its B-configuration used for the radio image we presented
in Erlund et al (2007).   It is clear from figure 4 of that paper that
the VLA detects far more of the smooth extended hotspot emission than
MERLIN does, which tells us that the only compact structure within the
hotspot is the slender southern-most arc shown as blue contours in
Fig.\,\ref{fig:newchanmerl}. 

We suggest that the diffuse radio emission detected by the VLA, that
MERLIN screens out, arises from adiabatic expansion of plasma emerging
from the north X-ray peak, thought to be synchrotron.  This flow then
ultimately shocks on the inter-galactic medium at the southern-most
point of this source, revealed by the slender MERLIN arc.

The considerable expansion losses out of this region will inevitably
shift any spectral features, such as a turnover at low Lorentz factors
still lower.  The only information that these observations give about
such a turnover in this particular object is that following these
losses, though reacceleration at higher energies at the southern-most
arc-shaped shock is taking place, there is evidence of ICCMB at
infra-red wavelengths that would be arising from electrons with
Lorentz factors of $\sim 24$.  We draw the readers attention to
  Figure\,\ref{fig:ssed} which illustrates the likely superposition of
the different emission processes.  

\subsection{Constraints imposed by the distance between the hotspot
  components}
\label{sec:constraints}

\begin{figure*}
  \center \rotatebox{0}{ \resizebox{!}{9cm}
    {\includegraphics{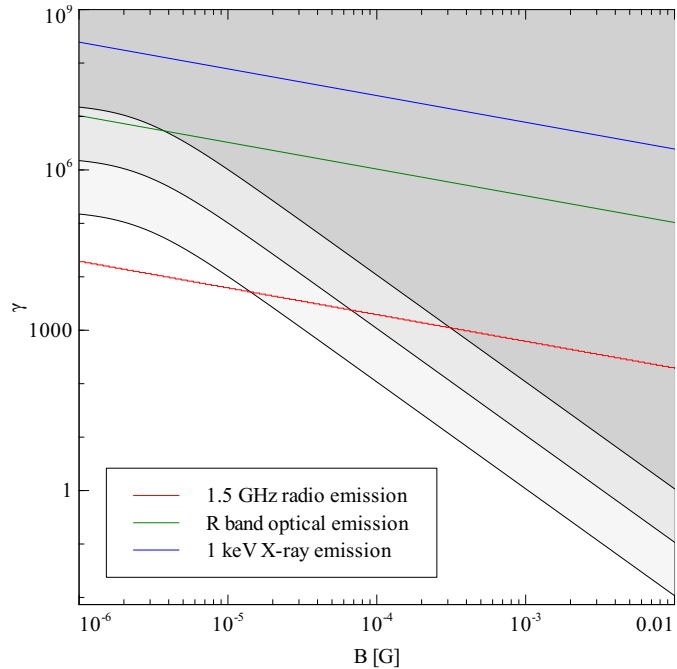}} }
  \caption[Constraints from the distance between the hotspot
  components]{The black lines represent the dependence of the Lorentz
    factor on magnetic field strength (from Equation \ref{equ:gbcool})
    imposing a cooling time equivalent to the time taken for the
    electrons to travel between the north X-ray peak and south radio
    hotspot.  The value of $d_{\rm max}$ used in Equation
    \ref{equ:gbcool} is $27$\kpc, the deprojected distance between the
    X-ray and radio peak assuming an orientation of $45\deg$ to the
    line of sight. This is a lower limit, as 4C\,74.26 is a quasar.
    The lines from top to bottom represent jet speeds after the inner
    shock of $\beta = 1, 0.1$ and $0.01$.  The red, green and blue
    lines represent the relationship between Lorentz factor and
    magnetic field strength responsible for producing 1.5\ghz, R band
    and 1\kev\ emission.}
\label{fig:gbreak}
\end{figure*}

If there is no second shock, then high Lorentz factor electrons which
are produced in a shock at the bright north X-ray peak, must have time
to travel to the radio peak without cooling.  The distance that
electrons can travel before cooling, $d_{\rm max}$, depends on the
speed of the jet after the initial shock, $\beta c$, and the time it
takes for an electron to lose its energy, $\mathcal{T}_{\rm loss}$.
\begin{equation}
  d_{\rm max} = c \beta \mathcal{T}_{\rm loss}
\label{equ:d}
\end{equation}
where $c$ is the speed of light and $\beta$ is the speed of the jet in
units of $c$.  If the flow was not expanding, then the particle energy
losses would be due to inverse-Compton and synchrotron losses only.
\begin{equation}
  \frac{1}{\mathcal{T}_{\rm loss}} = \frac{1}{\mathcal{T}_{\rm IC}} + \frac{1}{\mathcal{T}_{\rm sync}}
\end{equation}
These timescales (added in this fashion so that the shortest cooling
time dominates the loss timescale) represent the time taken for an
electron to cool via inverse-Compton scattering (the inverse-Compton
cooling timescale, $\mathcal{T}_{\rm IC}$) and synchrotron emission
(the synchrotron cooling timescale, $\mathcal{T}_{\rm sync}$).  The
inverse-Compton cooling timescale is given by
\begin{equation}
\mathcal{T}_{\rm IC} \sim\frac{3}{4}\frac{m_{\rm e}c}{\sigma_{\rm T}}\frac{1}{\mathcal{U}_{\rm ph}\gamma}
\label{equ:tic}
\end{equation}
where $m_{\rm e}$ is the mass of an electron, $c$ is the speed of
light, $\sigma_{\rm T}$ is the Thomson cross-section and
$\mathcal{U}_{\rm ph}$ is the energy density of the photons being
up-scattered.  $\gamma$ is the Lorentz factor of the electrons
responsible for the emission. If we consider synchrotron emission as
the up-scattering of virtual magnetic photons then we get a similar
expression for the synchrotron cooling time, $\mathcal{T}_{\rm sync}$,
\begin{equation}
\mathcal{T}_{\rm sync} \sim\frac{3}{4}\frac{m_{\rm e}c}{\sigma_{\rm T}}\frac{1}{\mathcal{U}_{\rm B}\gamma}
\label{equ:tsync}
\end{equation}
where $\mathcal{U}_{\rm B}$ is the energy density in the magnetic
field in c.g.s. units. Both timescales are in seconds.
Therefore the losses only depend on the energy density of the magnetic
field and photon fields available to be up-scattered.  Assuming that
the only important photon field is the cosmic microwave background
(i.e. $\mathcal{U}_{\rm ph} = \mathcal{U}_{\rm CMB}$) and substituting
this into Equation \ref{equ:d} gives
\begin{equation}
  \gamma = \frac{cu}{d_{\rm max}}\frac{3 m_{\rm e} c}{4 \sigma_{\rm T}}\left[\frac{B^2}{8\pi}+ \mathcal{U}_{\rm CMB}\right]^{-1}
\label{equ:gbcool}
\end{equation}
where $\mathcal{U}_{\rm CMB}$ is the energy density of the CMB at the
redshift of 4C\,74.26. Fig. \ref{fig:gbreak} shows that electrons with
radio-emitting Lorentz factors can survive to the radio hotspot, but
that the optical- and X-ray-emitting ones cannot.  Note that this is
an upper-limit as adiabatic losses would reduce the distance that
particles can travel without cooling, as would other inverse Compton
losses such as synchrotron self Compton (SSC) emission.  A second
shock is therefore required to re-accelerate particles further. 


\subsection{Consequences for models of the southern hotspot complex of
 4C\,74.26}

The initial observations and analysis of the southern hotspot complex
of 4C\,74.26 (see \citealt{erlund07}) suggested that three models in
the literature could reproduce the observed X-ray / radio emission:
the dentist drill's model \citep{scheuer82}, beaming from a
relativistic decelerating flow \citep{georganopouloskazanas04} and a
simple spine / sheath model \citep{chiaberge00}.

However, the spatially-resolved multi-wavelength morphology of the
hotspot complex makes it hard to see how a simple sheath model
\citep{chiaberge00} could produce its different spatial and broad-band
spectral characteristics.

The success of relativistic decelerating jet model put forward by
\citet{georganopouloskazanas04} is based on the idea that the initial
X-ray peak fades gradually towards the radio emission.  The new
\chandra\ observations rule out this model linking the X-ray peak to
the radio hotspot as they are clearly detected as two distinct X-ray
features (see Figs \ref{fig:nsxprof} and \ref{fig:nprofile} as well as
the image in Fig. \ref{fig:newchanmerl}), and not a gradual change.

\citet{scheuer82} dentist's drill model, if applicable to the
  southern hotspot of 4C74.26, suggests that the X-ray peak is due to
  synchrotron emission.  Constraints on the radiative lifetimes of
  synchrotron-emitting particles rule out all models requiring
  high-energy particles to be transported from the X-ray peak to the
  arc. The dentist's drill model implies that the jet is moving,
  possibly near the core or due to some instability.  This causes an
  offset between the current location of the jet terminus and where it
  used to be.  While we cannot completely rule out this model, it
  seems unlikely on the basis of the timescales implied by any
  synchrotron contribution from the optical/IR emission from the
  southern arc.

Another model, not considered in the analysis of the initial,
low-resolution data on 4C\,74.26, is that of a double
shock. Separated, high-frequency emitting shocks are seen in
supernova.  The outer shock is with the ISM and the inner (reverse)
shock is where ejecta ram into the slowed, post-shock material
\citep{reynolds08}.  High spatial resolution radio monitoring
observations of the cores of quasars show radio emitting blobs of
plasma being ejected \citep{kellermann04,listerhoman05}.  It is
reasonable to suppose that at the terminus of the jet, we could have
the same effect as in supernovae, with an outer shock, where the IGM
interacts with the advancing lobe head, and an inner shock, where the
blobs of plasma which have travelled down the jet collide with the
slowed post-shock plasma (e.g. \citealt{meisenheimer89}).  It is not
surprising that, in the case of a jet terminus, the inner shock is
planar and the outer shock more laterally extended and arc-like.  A
region of high pressure will exist between the two shocks which could
account for the diffuse X-ray, optical and radio emission detected
there.  The strongest indication that what we are seeing is
fundamental to shock structure at the jet terminus is the growing
number of radio quasars, for which 4C\,74.26 is as a prototype, that
show similar (if less clearly resolved) phenomena, in the same sense.
For example, 4C\,19.44 \citep{sambruna04}, as well as PKS\,1055$+$201
which shows two X-ray sources at the jet terminus, one inset from the
radio peak along the line of the jet, which is clearly detected in
this source \citep{schwartz06}.




\begin{table*} 
\center
\begin{tabular}{lllllll}
\hline
Region     & statistic                     & bin & $\Gamma$       & Norm                      & $F_{\rm X }$             & $L_{\rm X}$  \\
           &                               &     &                & $10^{-6}$\photpkevpcmsqps\ & $10^{-14}$\ergpcmsqps\  & $10^{42}$\ergps\  \\
\hline
northern & $\chi^2_{\nu} = 1.0$ [$10$]    & 20  & $1.7\pm 0.2$   & $9.3\pm 0.9$              & $3.3^{+0.5}_{-0.6}$      & $0.93^{+0.21}_{-0.15}$ \\
X-ray peak & $\text{cstat} = 649$ [$1272$] & 0   & $1.7\pm 0.1$   & $9.3^{+0.9}_{-0.8}$         & $3.3\pm 0.4$           & $1.0\pm 0.1$ \\
\hline
southern 
X-ray arc  & $\text{cstat} = 457$ [$1272$] & 0   & $2.1\pm 0.2$  & $5.0^{+0.7}_{-0.6}$          & $1.1\pm 0.2$            & $0.32\pm 0.07$\\
\hline
hotspot complex & $\chi^2_{\nu} = 1.1$ [$17$] & 20 & $1.7\pm 0.1$  & $13\pm 1$                 & $4.7^{+0.5}_{-0.6}$       & $1.2\pm 0.2$ \\
\hline
\end{tabular}
\caption[Spectral fits of the X-ray components of 4C\,74.26 using \chandra\ data]
{Spectral fits are in the 0.5--7\,\kev\ energy band in order to ensure the different regions are fitted over comparable ranges. 
  The columns are: (1) the region analysed,
  (2) statistic used for the analysis with the best-fit value and, in square brackets, the number of degrees of freedom 
  or number of spectral bins, in the case of C-statistics, (3) the number of photons binned together, 
  (4) $\Gamma$ is the X-ray photon index, (5) Norm is the normalisation of the power-law, 
  (6) $F_{\rm X}$ is the observed X-ray flux in the rest-frame $2-10$\kev\ band,
  (7) $L_{\rm X}$ is the absorption corrected X-ray luminosity in the rest-frame $2-10$\kev\ band.}
\label{table:newchanspec} 
\end{table*}

\section{Conclusions}

4C\,74.26 is the largest known radio quasar and is one of the largest
radio sources in the Local Universe, spanning 1.1\mpc\ on the sky (in
projected distance).  Its southern hotspot complex is detected at
radio, infrared, optical and X-ray wavelengths.  The X-ray components
of the southern hotspot complex are composed of a north X-ray bright
peak (the most luminous X-ray component of this radio quasar, ignoring
its nucleus) and a south X-ray arc.

The north X-ray peak is not detected in our observations at other
wavelengths, but is thought to be synchrotron emission arising at a
shock, perpendicular to the axis of the jet.  The lateral extent of
the X-ray peak constrains the opening angle of the jet to be $\sim
0.5\deg$ if the jet axis is $45\deg$ to the line of sight.

Considerable diffuse radio emission is associated with the downstream
region of this shock, consistent with expansion losses out of the
first, X-ray-luminous, shock.

There is a considerable physical offset between this X-ray peak and a
second shock structure is also detected as an arc shape: 10\as\
(19\kpc\ projected onto the sky) downstream.  This is clearly
delineated in radio synchrotron revealed by MERLIN, and has co-spatial
infra-red, optical and X-ray emission.  The X-ray emission is
inconsistent with being synchrotron emission, but can be explained as
ICCMB if the thin arc region of the shock has a magnetic field
strength $\sim 4$\,\% of the minimum energy value.

These two very different shocks, and the offset between them, is
clearly discerned in this object because it is unusually close and
unusually large.  We note that difficulties in interpretation of X-ray
hotspots in the general case for objects that are both smaller and
further away will be compounded because resolution insufficient to
separate X-ray and radio emission that is actually physically offset
and having different physical origins. 

In addition, the angular separation of the spatially-extended double
shock structure from the active nucleus that drives this by $\sim
550\,$kpc could present a challenge for connecting ``unidentified''
hard X-ray or Fermi sources with their origins.  Depending on how high
the Lorentz factors are to which protons are accelerated, it is
possible that such primary acceleration sites within hotspots could be
Fermi detections: for comparision the Fermi sensitivity after one year
is $10^{-12} {\rm erg\,cm^{-2}\,s{-1}}$ (over 4 decades of frequency,
$10^{22}$--$10^{24}$\,Hz) is slightly over an order of magnitude more
flux density than we detect in the much smaller-bandwidth Chandra band 
$8 \times 10^{-14} {\rm erg\,cm^{-2}\,s{-1}}$ around $10^{17}$\,Hz.  

\section*{Acknowledgements}

MCE acknowledges STFC for financial support. ACF and KMB thank the
Royal Society.  Based in part on observations obtained at the Gemini
Observatory, which is operated by the Association of Universities for
Research in Astronomy, Inc., under a cooperative agreement with the
NSF on behalf of the Gemini partnership: the National Science
Foundation (United States), the Science and Technology Facilities
Council (United Kingdom), the National Research Council (Canada),
CONICYT (Chile), the Australian Research Council (Australia),
Ministério da Ciência e Tecnologia (Brazil) and SECYT (Argentina).
MERLIN is a National Facility operated by the University of Manchester
at Jodrell Bank Observatory on behalf of STFC, formerly known as
PPARC.

\bibliographystyle{mn2e} 

\end{document}

%% file: rgb.tex
\definecolor{AliceBlue}{rgb}{0.94,0.97,1.00}
\definecolor{AntiqueWhite1}{rgb}{1.00,0.94,0.86}
\definecolor{AntiqueWhite2}{rgb}{0.93,0.87,0.80}
\definecolor{AntiqueWhite3}{rgb}{0.80,0.75,0.69}
\definecolor{AntiqueWhite4}{rgb}{0.55,0.51,0.47}
\definecolor{AntiqueWhite}{rgb}{0.98,0.92,0.84}
\definecolor{BlanchedAlmond}{rgb}{1.00,0.92,0.80}
\definecolor{BlueViolet}{rgb}{0.54,0.17,0.89}
\definecolor{CadetBlue1}{rgb}{0.60,0.96,1.00}
\definecolor{CadetBlue2}{rgb}{0.56,0.90,0.93}
\definecolor{CadetBlue3}{rgb}{0.48,0.77,0.80}
\definecolor{CadetBlue4}{rgb}{0.33,0.53,0.55}
\definecolor{CadetBlue}{rgb}{0.37,0.62,0.63}
\definecolor{CornflowerBlue}{rgb}{0.39,0.58,0.93}
\definecolor{DarkBlue}{rgb}{0.00,0.00,0.55}
\definecolor{DarkCyan}{rgb}{0.00,0.55,0.55}
\definecolor{DarkGoldenrod1}{rgb}{1.00,0.73,0.06}
\definecolor{DarkGoldenrod2}{rgb}{0.93,0.68,0.05}
\definecolor{DarkGoldenrod3}{rgb}{0.80,0.58,0.05}
\definecolor{DarkGoldenrod4}{rgb}{0.55,0.40,0.03}
\definecolor{DarkGoldenrod}{rgb}{0.72,0.53,0.04}
\definecolor{DarkGray}{rgb}{0.66,0.66,0.66}
\definecolor{DarkGreen}{rgb}{0.00,0.39,0.00}
\definecolor{DarkGrey}{rgb}{0.66,0.66,0.66}
\definecolor{DarkKhaki}{rgb}{0.74,0.72,0.42}
\definecolor{DarkMagenta}{rgb}{0.55,0.00,0.55}
\definecolor{DarkOliveGreen1}{rgb}{0.79,1.00,0.44}
\definecolor{DarkOliveGreen2}{rgb}{0.74,0.93,0.41}
\definecolor{DarkOliveGreen3}{rgb}{0.64,0.80,0.35}
\definecolor{DarkOliveGreen4}{rgb}{0.43,0.55,0.24}
\definecolor{DarkOliveGreen}{rgb}{0.33,0.42,0.18}
\definecolor{DarkOrange1}{rgb}{1.00,0.50,0.00}
\definecolor{DarkOrange2}{rgb}{0.93,0.46,0.00}
\definecolor{DarkOrange3}{rgb}{0.80,0.40,0.00}
\definecolor{DarkOrange4}{rgb}{0.55,0.27,0.00}
\definecolor{DarkOrange}{rgb}{1.00,0.55,0.00}
\definecolor{DarkOrchid1}{rgb}{0.75,0.24,1.00}
\definecolor{DarkOrchid2}{rgb}{0.70,0.23,0.93}
\definecolor{DarkOrchid3}{rgb}{0.60,0.20,0.80}
\definecolor{DarkOrchid4}{rgb}{0.41,0.13,0.55}
\definecolor{DarkOrchid}{rgb}{0.60,0.20,0.80}
\definecolor{DarkRed}{rgb}{0.55,0.00,0.00}
\definecolor{DarkSalmon}{rgb}{0.91,0.59,0.48}
\definecolor{DarkSeaGreen1}{rgb}{0.76,1.00,0.76}
\definecolor{DarkSeaGreen2}{rgb}{0.71,0.93,0.71}
\definecolor{DarkSeaGreen3}{rgb}{0.61,0.80,0.61}
\definecolor{DarkSeaGreen4}{rgb}{0.41,0.55,0.41}
\definecolor{DarkSeaGreen}{rgb}{0.56,0.74,0.56}
\definecolor{DarkSlateBlue}{rgb}{0.28,0.24,0.55}
\definecolor{DarkSlateGray1}{rgb}{0.59,1.00,1.00}
\definecolor{DarkSlateGray2}{rgb}{0.55,0.93,0.93}
\definecolor{DarkSlateGray3}{rgb}{0.47,0.80,0.80}
\definecolor{DarkSlateGray4}{rgb}{0.32,0.55,0.55}
\definecolor{DarkSlateGray}{rgb}{0.18,0.31,0.31}
\definecolor{DarkSlateGrey}{rgb}{0.18,0.31,0.31}
\definecolor{DarkTurquoise}{rgb}{0.00,0.81,0.82}
\definecolor{DarkViolet}{rgb}{0.58,0.00,0.83}
\definecolor{DeepPink1}{rgb}{1.00,0.08,0.58}
\definecolor{DeepPink2}{rgb}{0.93,0.07,0.54}
\definecolor{DeepPink3}{rgb}{0.80,0.06,0.46}
\definecolor{DeepPink4}{rgb}{0.55,0.04,0.31}
\definecolor{DeepPink}{rgb}{1.00,0.08,0.58}
\definecolor{DeepSkyBlue1}{rgb}{0.00,0.75,1.00}
\definecolor{DeepSkyBlue2}{rgb}{0.00,0.70,0.93}
\definecolor{DeepSkyBlue3}{rgb}{0.00,0.60,0.80}
\definecolor{DeepSkyBlue4}{rgb}{0.00,0.41,0.55}
\definecolor{DeepSkyBlue}{rgb}{0.00,0.75,1.00}
\definecolor{DimGray}{rgb}{0.41,0.41,0.41}
\definecolor{DimGrey}{rgb}{0.41,0.41,0.41}
\definecolor{DodgerBlue1}{rgb}{0.12,0.56,1.00}
\definecolor{DodgerBlue2}{rgb}{0.11,0.53,0.93}
\definecolor{DodgerBlue3}{rgb}{0.09,0.45,0.80}
\definecolor{DodgerBlue4}{rgb}{0.06,0.31,0.55}
\definecolor{DodgerBlue}{rgb}{0.12,0.56,1.00}
\definecolor{FloralWhite}{rgb}{1.00,0.98,0.94}
\definecolor{ForestGreen}{rgb}{0.13,0.55,0.13}
\definecolor{GhostWhite}{rgb}{0.97,0.97,1.00}
\definecolor{GreenYellow}{rgb}{0.68,1.00,0.18}
\definecolor{HotPink1}{rgb}{1.00,0.43,0.71}
\definecolor{HotPink2}{rgb}{0.93,0.42,0.65}
\definecolor{HotPink3}{rgb}{0.80,0.38,0.56}
\definecolor{HotPink4}{rgb}{0.55,0.23,0.38}
\definecolor{HotPink}{rgb}{1.00,0.41,0.71}
\definecolor{IndianRed1}{rgb}{1.00,0.42,0.42}
\definecolor{IndianRed2}{rgb}{0.93,0.39,0.39}
\definecolor{IndianRed3}{rgb}{0.80,0.33,0.33}
\definecolor{IndianRed4}{rgb}{0.55,0.23,0.23}
\definecolor{IndianRed}{rgb}{0.80,0.36,0.36}
\definecolor{LavenderBlush1}{rgb}{1.00,0.94,0.96}
\definecolor{LavenderBlush2}{rgb}{0.93,0.88,0.90}
\definecolor{LavenderBlush3}{rgb}{0.80,0.76,0.77}
\definecolor{LavenderBlush4}{rgb}{0.55,0.51,0.53}
\definecolor{LavenderBlush}{rgb}{1.00,0.94,0.96}
\definecolor{LawnGreen}{rgb}{0.49,0.99,0.00}
\definecolor{LemonChiffon1}{rgb}{1.00,0.98,0.80}
\definecolor{LemonChiffon2}{rgb}{0.93,0.91,0.75}
\definecolor{LemonChiffon3}{rgb}{0.80,0.79,0.65}
\definecolor{LemonChiffon4}{rgb}{0.55,0.54,0.44}
\definecolor{LemonChiffon}{rgb}{1.00,0.98,0.80}
\definecolor{LightBlue1}{rgb}{0.75,0.94,1.00}
\definecolor{LightBlue2}{rgb}{0.70,0.87,0.93}
\definecolor{LightBlue3}{rgb}{0.60,0.75,0.80}
\definecolor{LightBlue4}{rgb}{0.41,0.51,0.55}
\definecolor{LightBlue}{rgb}{0.68,0.85,0.90}
\definecolor{LightCoral}{rgb}{0.94,0.50,0.50}
\definecolor{LightCyan1}{rgb}{0.88,1.00,1.00}
\definecolor{LightCyan2}{rgb}{0.82,0.93,0.93}
\definecolor{LightCyan3}{rgb}{0.71,0.80,0.80}
\definecolor{LightCyan4}{rgb}{0.48,0.55,0.55}
\definecolor{LightCyan}{rgb}{0.88,1.00,1.00}
\definecolor{LightGoldenrod1}{rgb}{1.00,0.93,0.55}
\definecolor{LightGoldenrod2}{rgb}{0.93,0.86,0.51}
\definecolor{LightGoldenrod3}{rgb}{0.80,0.75,0.44}
\definecolor{LightGoldenrod4}{rgb}{0.55,0.51,0.30}
\definecolor{LightGoldenrodYellow}{rgb}{0.98,0.98,0.82}
\definecolor{LightGoldenrod}{rgb}{0.93,0.87,0.51}
\definecolor{LightGray}{rgb}{0.83,0.83,0.83}
\definecolor{LightGreen}{rgb}{0.56,0.93,0.56}
\definecolor{LightGrey}{rgb}{0.83,0.83,0.83}
\definecolor{LightPink1}{rgb}{1.00,0.68,0.73}
\definecolor{LightPink2}{rgb}{0.93,0.64,0.68}
\definecolor{LightPink3}{rgb}{0.80,0.55,0.58}
\definecolor{LightPink4}{rgb}{0.55,0.37,0.40}
\definecolor{LightPink}{rgb}{1.00,0.71,0.76}
\definecolor{LightSalmon1}{rgb}{1.00,0.63,0.48}
\definecolor{LightSalmon2}{rgb}{0.93,0.58,0.45}
\definecolor{LightSalmon3}{rgb}{0.80,0.51,0.38}
\definecolor{LightSalmon4}{rgb}{0.55,0.34,0.26}
\definecolor{LightSalmon}{rgb}{1.00,0.63,0.48}
\definecolor{LightSeaGreen}{rgb}{0.13,0.70,0.67}
\definecolor{LightSkyBlue1}{rgb}{0.69,0.89,1.00}
\definecolor{LightSkyBlue2}{rgb}{0.64,0.83,0.93}
\definecolor{LightSkyBlue3}{rgb}{0.55,0.71,0.80}
\definecolor{LightSkyBlue4}{rgb}{0.38,0.48,0.55}
\definecolor{LightSkyBlue}{rgb}{0.53,0.81,0.98}
\definecolor{LightSlateBlue}{rgb}{0.52,0.44,1.00}
\definecolor{LightSlateGray}{rgb}{0.47,0.53,0.60}
\definecolor{LightSlateGrey}{rgb}{0.47,0.53,0.60}
\definecolor{LightSteelBlue1}{rgb}{0.79,0.88,1.00}
\definecolor{LightSteelBlue2}{rgb}{0.74,0.82,0.93}
\definecolor{LightSteelBlue3}{rgb}{0.64,0.71,0.80}
\definecolor{LightSteelBlue4}{rgb}{0.43,0.48,0.55}
\definecolor{LightSteelBlue}{rgb}{0.69,0.77,0.87}
\definecolor{LightYellow1}{rgb}{1.00,1.00,0.88}
\definecolor{LightYellow2}{rgb}{0.93,0.93,0.82}
\definecolor{LightYellow3}{rgb}{0.80,0.80,0.71}
\definecolor{LightYellow4}{rgb}{0.55,0.55,0.48}
\definecolor{LightYellow}{rgb}{1.00,1.00,0.88}
\definecolor{LimeGreen}{rgb}{0.20,0.80,0.20}
\definecolor{MediumAquamarine}{rgb}{0.40,0.80,0.67}
\definecolor{MediumBlue}{rgb}{0.00,0.00,0.80}
\definecolor{MediumOrchid1}{rgb}{0.88,0.40,1.00}
\definecolor{MediumOrchid2}{rgb}{0.82,0.37,0.93}
\definecolor{MediumOrchid3}{rgb}{0.71,0.32,0.80}
\definecolor{MediumOrchid4}{rgb}{0.48,0.22,0.55}
\definecolor{MediumOrchid}{rgb}{0.73,0.33,0.83}
\definecolor{MediumPurple1}{rgb}{0.67,0.51,1.00}
\definecolor{MediumPurple2}{rgb}{0.62,0.47,0.93}
\definecolor{MediumPurple3}{rgb}{0.54,0.41,0.80}
\definecolor{MediumPurple4}{rgb}{0.36,0.28,0.55}
\definecolor{MediumPurple}{rgb}{0.58,0.44,0.86}
\definecolor{MediumSeaGreen}{rgb}{0.24,0.70,0.44}
\definecolor{MediumSlateBlue}{rgb}{0.48,0.41,0.93}
\definecolor{MediumSpringGreen}{rgb}{0.00,0.98,0.60}
\definecolor{MediumTurquoise}{rgb}{0.28,0.82,0.80}
\definecolor{MediumVioletRed}{rgb}{0.78,0.08,0.52}
\definecolor{MidnightBlue}{rgb}{0.10,0.10,0.44}
\definecolor{MintCream}{rgb}{0.96,1.00,0.98}
\definecolor{MistyRose1}{rgb}{1.00,0.89,0.88}
\definecolor{MistyRose2}{rgb}{0.93,0.84,0.82}
\definecolor{MistyRose3}{rgb}{0.80,0.72,0.71}
\definecolor{MistyRose4}{rgb}{0.55,0.49,0.48}
\definecolor{MistyRose}{rgb}{1.00,0.89,0.88}
\definecolor{NavajoWhite1}{rgb}{1.00,0.87,0.68}
\definecolor{NavajoWhite2}{rgb}{0.93,0.81,0.63}
\definecolor{NavajoWhite3}{rgb}{0.80,0.70,0.55}
\definecolor{NavajoWhite4}{rgb}{0.55,0.47,0.37}
\definecolor{NavajoWhite}{rgb}{1.00,0.87,0.68}
\definecolor{NavyBlue}{rgb}{0.00,0.00,0.50}
\definecolor{OldLace}{rgb}{0.99,0.96,0.90}
\definecolor{OliveDrab1}{rgb}{0.75,1.00,0.24}
\definecolor{OliveDrab2}{rgb}{0.70,0.93,0.23}
\definecolor{OliveDrab3}{rgb}{0.60,0.80,0.20}
\definecolor{OliveDrab4}{rgb}{0.41,0.55,0.13}
\definecolor{OliveDrab}{rgb}{0.42,0.56,0.14}
\definecolor{OrangeRed1}{rgb}{1.00,0.27,0.00}
\definecolor{OrangeRed2}{rgb}{0.93,0.25,0.00}
\definecolor{OrangeRed3}{rgb}{0.80,0.22,0.00}
\definecolor{OrangeRed4}{rgb}{0.55,0.15,0.00}
\definecolor{OrangeRed}{rgb}{1.00,0.27,0.00}
\definecolor{PaleGoldenrod}{rgb}{0.93,0.91,0.67}
\definecolor{PaleGreen1}{rgb}{0.60,1.00,0.60}
\definecolor{PaleGreen2}{rgb}{0.56,0.93,0.56}
\definecolor{PaleGreen3}{rgb}{0.49,0.80,0.49}
\definecolor{PaleGreen4}{rgb}{0.33,0.55,0.33}
\definecolor{PaleGreen}{rgb}{0.60,0.98,0.60}
\definecolor{PaleTurquoise1}{rgb}{0.73,1.00,1.00}
\definecolor{PaleTurquoise2}{rgb}{0.68,0.93,0.93}
\definecolor{PaleTurquoise3}{rgb}{0.59,0.80,0.80}
\definecolor{PaleTurquoise4}{rgb}{0.40,0.55,0.55}
\definecolor{PaleTurquoise}{rgb}{0.69,0.93,0.93}
\definecolor{PaleVioletRed1}{rgb}{1.00,0.51,0.67}
\definecolor{PaleVioletRed2}{rgb}{0.93,0.47,0.62}
\definecolor{PaleVioletRed3}{rgb}{0.80,0.41,0.54}
\definecolor{PaleVioletRed4}{rgb}{0.55,0.28,0.36}
\definecolor{PaleVioletRed}{rgb}{0.86,0.44,0.58}
\definecolor{PapayaWhip}{rgb}{1.00,0.94,0.84}
\definecolor{PeachPuff1}{rgb}{1.00,0.85,0.73}
\definecolor{PeachPuff2}{rgb}{0.93,0.80,0.68}
\definecolor{PeachPuff3}{rgb}{0.80,0.69,0.58}
\definecolor{PeachPuff4}{rgb}{0.55,0.47,0.40}
\definecolor{PeachPuff}{rgb}{1.00,0.85,0.73}
\definecolor{PowderBlue}{rgb}{0.69,0.88,0.90}
\definecolor{RosyBrown1}{rgb}{1.00,0.76,0.76}
\definecolor{RosyBrown2}{rgb}{0.93,0.71,0.71}
\definecolor{RosyBrown3}{rgb}{0.80,0.61,0.61}
\definecolor{RosyBrown4}{rgb}{0.55,0.41,0.41}
\definecolor{RosyBrown}{rgb}{0.74,0.56,0.56}
\definecolor{RoyalBlue1}{rgb}{0.28,0.46,1.00}
\definecolor{RoyalBlue2}{rgb}{0.26,0.43,0.93}
\definecolor{RoyalBlue3}{rgb}{0.23,0.37,0.80}
\definecolor{RoyalBlue4}{rgb}{0.15,0.25,0.55}
\definecolor{RoyalBlue}{rgb}{0.25,0.41,0.88}
\definecolor{SaddleBrown}{rgb}{0.55,0.27,0.07}
\definecolor{SandyBrown}{rgb}{0.96,0.64,0.38}
\definecolor{SeaGreen1}{rgb}{0.33,1.00,0.62}
\definecolor{SeaGreen2}{rgb}{0.31,0.93,0.58}
\definecolor{SeaGreen3}{rgb}{0.26,0.80,0.50}
\definecolor{SeaGreen4}{rgb}{0.18,0.55,0.34}
\definecolor{SeaGreen}{rgb}{0.18,0.55,0.34}
\definecolor{SkyBlue1}{rgb}{0.53,0.81,1.00}
\definecolor{SkyBlue2}{rgb}{0.49,0.75,0.93}
\definecolor{SkyBlue3}{rgb}{0.42,0.65,0.80}
\definecolor{SkyBlue4}{rgb}{0.29,0.44,0.55}
\definecolor{SkyBlue}{rgb}{0.53,0.81,0.92}
\definecolor{SlateBlue1}{rgb}{0.51,0.44,1.00}
\definecolor{SlateBlue2}{rgb}{0.48,0.40,0.93}
\definecolor{SlateBlue3}{rgb}{0.41,0.35,0.80}
\definecolor{SlateBlue4}{rgb}{0.28,0.24,0.55}
\definecolor{SlateBlue}{rgb}{0.42,0.35,0.80}
\definecolor{SlateGray1}{rgb}{0.78,0.89,1.00}
\definecolor{SlateGray2}{rgb}{0.73,0.83,0.93}
\definecolor{SlateGray3}{rgb}{0.62,0.71,0.80}
\definecolor{SlateGray4}{rgb}{0.42,0.48,0.55}
\definecolor{SlateGray}{rgb}{0.44,0.50,0.56}
\definecolor{SlateGrey}{rgb}{0.44,0.50,0.56}
\definecolor{SpringGreen1}{rgb}{0.00,1.00,0.50}
\definecolor{SpringGreen2}{rgb}{0.00,0.93,0.46}
\definecolor{SpringGreen3}{rgb}{0.00,0.80,0.40}
\definecolor{SpringGreen4}{rgb}{0.00,0.55,0.27}
\definecolor{SpringGreen}{rgb}{0.00,1.00,0.50}
\definecolor{SteelBlue1}{rgb}{0.39,0.72,1.00}
\definecolor{SteelBlue2}{rgb}{0.36,0.67,0.93}
\definecolor{SteelBlue3}{rgb}{0.31,0.58,0.80}
\definecolor{SteelBlue4}{rgb}{0.21,0.39,0.55}
\definecolor{SteelBlue}{rgb}{0.27,0.51,0.71}
\definecolor{VioletRed1}{rgb}{1.00,0.24,0.59}
\definecolor{VioletRed2}{rgb}{0.93,0.23,0.55}
\definecolor{VioletRed3}{rgb}{0.80,0.20,0.47}
\definecolor{VioletRed4}{rgb}{0.55,0.13,0.32}
\definecolor{VioletRed}{rgb}{0.82,0.13,0.56}
\definecolor{WhiteSmoke}{rgb}{0.96,0.96,0.96}
\definecolor{YellowGreen}{rgb}{0.60,0.80,0.20}
\definecolor{aliceblue}{rgb}{0.94,0.97,1.00}
\definecolor{antiquewhite}{rgb}{0.98,0.92,0.84}
\definecolor{aquamarine1}{rgb}{0.50,1.00,0.83}
\definecolor{aquamarine2}{rgb}{0.46,0.93,0.78}
\definecolor{aquamarine3}{rgb}{0.40,0.80,0.67}
\definecolor{aquamarine4}{rgb}{0.27,0.55,0.45}
\definecolor{aquamarine}{rgb}{0.50,1.00,0.83}
\definecolor{azure1}{rgb}{0.94,1.00,1.00}
\definecolor{azure2}{rgb}{0.88,0.93,0.93}
\definecolor{azure3}{rgb}{0.76,0.80,0.80}
\definecolor{azure4}{rgb}{0.51,0.55,0.55}
\definecolor{azure}{rgb}{0.94,1.00,1.00}
\definecolor{beige}{rgb}{0.96,0.96,0.86}
\definecolor{bisque1}{rgb}{1.00,0.89,0.77}
\definecolor{bisque2}{rgb}{0.93,0.84,0.72}
\definecolor{bisque3}{rgb}{0.80,0.72,0.62}
\definecolor{bisque4}{rgb}{0.55,0.49,0.42}
\definecolor{bisque}{rgb}{1.00,0.89,0.77}
\definecolor{black}{rgb}{0.00,0.00,0.00}
\definecolor{blanchedalmond}{rgb}{1.00,0.92,0.80}
\definecolor{blue1}{rgb}{0.00,0.00,1.00}
\definecolor{blue2}{rgb}{0.00,0.00,0.93}
\definecolor{blue3}{rgb}{0.00,0.00,0.80}
\definecolor{blue4}{rgb}{0.00,0.00,0.55}
\definecolor{blueviolet}{rgb}{0.54,0.17,0.89}
\definecolor{blue}{rgb}{0.00,0.00,1.00}
\definecolor{brown1}{rgb}{1.00,0.25,0.25}
\definecolor{brown2}{rgb}{0.93,0.23,0.23}
\definecolor{brown3}{rgb}{0.80,0.20,0.20}
\definecolor{brown4}{rgb}{0.55,0.14,0.14}
\definecolor{brown}{rgb}{0.65,0.16,0.16}
\definecolor{burlywood1}{rgb}{1.00,0.83,0.61}
\definecolor{burlywood2}{rgb}{0.93,0.77,0.57}
\definecolor{burlywood3}{rgb}{0.80,0.67,0.49}
\definecolor{burlywood4}{rgb}{0.55,0.45,0.33}
\definecolor{burlywood}{rgb}{0.87,0.72,0.53}
\definecolor{cadetblue}{rgb}{0.37,0.62,0.63}
\definecolor{chartreuse1}{rgb}{0.50,1.00,0.00}
\definecolor{chartreuse2}{rgb}{0.46,0.93,0.00}
\definecolor{chartreuse3}{rgb}{0.40,0.80,0.00}
\definecolor{chartreuse4}{rgb}{0.27,0.55,0.00}
\definecolor{chartreuse}{rgb}{0.50,1.00,0.00}
\definecolor{chocolate1}{rgb}{1.00,0.50,0.14}
\definecolor{chocolate2}{rgb}{0.93,0.46,0.13}
\definecolor{chocolate3}{rgb}{0.80,0.40,0.11}
\definecolor{chocolate4}{rgb}{0.55,0.27,0.07}
\definecolor{chocolate}{rgb}{0.82,0.41,0.12}
\definecolor{coral1}{rgb}{1.00,0.45,0.34}
\definecolor{coral2}{rgb}{0.93,0.42,0.31}
\definecolor{coral3}{rgb}{0.80,0.36,0.27}
\definecolor{coral4}{rgb}{0.55,0.24,0.18}
\definecolor{coral}{rgb}{1.00,0.50,0.31}
\definecolor{cornflowerblue}{rgb}{0.39,0.58,0.93}
\definecolor{cornsilk1}{rgb}{1.00,0.97,0.86}
\definecolor{cornsilk2}{rgb}{0.93,0.91,0.80}
\definecolor{cornsilk3}{rgb}{0.80,0.78,0.69}
\definecolor{cornsilk4}{rgb}{0.55,0.53,0.47}
\definecolor{cornsilk}{rgb}{1.00,0.97,0.86}
\definecolor{cyan1}{rgb}{0.00,1.00,1.00}
\definecolor{cyan2}{rgb}{0.00,0.93,0.93}
\definecolor{cyan3}{rgb}{0.00,0.80,0.80}
\definecolor{cyan4}{rgb}{0.00,0.55,0.55}
\definecolor{cyan}{rgb}{0.00,1.00,1.00}
\definecolor{darkblue}{rgb}{0.00,0.00,0.55}
\definecolor{darkcyan}{rgb}{0.00,0.55,0.55}
\definecolor{darkgoldenrod}{rgb}{0.72,0.53,0.04}
\definecolor{darkgray}{rgb}{0.66,0.66,0.66}
\definecolor{darkgreen}{rgb}{0.00,0.39,0.00}
\definecolor{darkgrey}{rgb}{0.66,0.66,0.66}
\definecolor{darkkhaki}{rgb}{0.74,0.72,0.42}
\definecolor{darkmagenta}{rgb}{0.55,0.00,0.55}
\definecolor{darkolive}{rgb}{0.33,0.42,0.18}
\definecolor{darkorange}{rgb}{1.00,0.55,0.00}
\definecolor{darkorchid}{rgb}{0.60,0.20,0.80}
\definecolor{darkred}{rgb}{0.55,0.00,0.00}
\definecolor{darksalmon}{rgb}{0.91,0.59,0.48}
\definecolor{darksea}{rgb}{0.56,0.74,0.56}
\definecolor{darkslate}{rgb}{0.18,0.31,0.31}
\definecolor{darkslate}{rgb}{0.18,0.31,0.31}
\definecolor{darkslate}{rgb}{0.28,0.24,0.55}
\definecolor{darkturquoise}{rgb}{0.00,0.81,0.82}
\definecolor{darkviolet}{rgb}{0.58,0.00,0.83}
\definecolor{deeppink}{rgb}{1.00,0.08,0.58}
\definecolor{deepsky}{rgb}{0.00,0.75,1.00}
\definecolor{dimgray}{rgb}{0.41,0.41,0.41}
\definecolor{dimgrey}{rgb}{0.41,0.41,0.41}
\definecolor{dodgerblue}{rgb}{0.12,0.56,1.00}
\definecolor{firebrick1}{rgb}{1.00,0.19,0.19}
\definecolor{firebrick2}{rgb}{0.93,0.17,0.17}
\definecolor{firebrick3}{rgb}{0.80,0.15,0.15}
\definecolor{firebrick4}{rgb}{0.55,0.10,0.10}
\definecolor{firebrick}{rgb}{0.70,0.13,0.13}
\definecolor{floralwhite}{rgb}{1.00,0.98,0.94}
\definecolor{forestgreen}{rgb}{0.13,0.55,0.13}
\definecolor{gainsboro}{rgb}{0.86,0.86,0.86}
\definecolor{ghostwhite}{rgb}{0.97,0.97,1.00}
\definecolor{gold1}{rgb}{1.00,0.84,0.00}
\definecolor{gold2}{rgb}{0.93,0.79,0.00}
\definecolor{gold3}{rgb}{0.80,0.68,0.00}
\definecolor{gold4}{rgb}{0.55,0.46,0.00}
\definecolor{goldenrod1}{rgb}{1.00,0.76,0.15}
\definecolor{goldenrod2}{rgb}{0.93,0.71,0.13}
\definecolor{goldenrod3}{rgb}{0.80,0.61,0.11}
\definecolor{goldenrod4}{rgb}{0.55,0.41,0.08}
\definecolor{goldenrod}{rgb}{0.85,0.65,0.13}
\definecolor{gold}{rgb}{1.00,0.84,0.00}
\definecolor{gray0}{rgb}{0.00,0.00,0.00}
\definecolor{gray100}{rgb}{1.00,1.00,1.00}
\definecolor{gray10}{rgb}{0.10,0.10,0.10}
\definecolor{gray11}{rgb}{0.11,0.11,0.11}
\definecolor{gray12}{rgb}{0.12,0.12,0.12}
\definecolor{gray13}{rgb}{0.13,0.13,0.13}
\definecolor{gray14}{rgb}{0.14,0.14,0.14}
\definecolor{gray15}{rgb}{0.15,0.15,0.15}
\definecolor{gray16}{rgb}{0.16,0.16,0.16}
\definecolor{gray17}{rgb}{0.17,0.17,0.17}
\definecolor{gray18}{rgb}{0.18,0.18,0.18}
\definecolor{gray19}{rgb}{0.19,0.19,0.19}
\definecolor{gray1}{rgb}{0.01,0.01,0.01}
\definecolor{gray20}{rgb}{0.20,0.20,0.20}
\definecolor{gray21}{rgb}{0.21,0.21,0.21}
\definecolor{gray22}{rgb}{0.22,0.22,0.22}
\definecolor{gray23}{rgb}{0.23,0.23,0.23}
\definecolor{gray24}{rgb}{0.24,0.24,0.24}
\definecolor{gray25}{rgb}{0.25,0.25,0.25}
\definecolor{gray26}{rgb}{0.26,0.26,0.26}
\definecolor{gray27}{rgb}{0.27,0.27,0.27}
\definecolor{gray28}{rgb}{0.28,0.28,0.28}
\definecolor{gray29}{rgb}{0.29,0.29,0.29}
\definecolor{gray2}{rgb}{0.02,0.02,0.02}
\definecolor{gray30}{rgb}{0.30,0.30,0.30}
\definecolor{gray31}{rgb}{0.31,0.31,0.31}
\definecolor{gray32}{rgb}{0.32,0.32,0.32}
\definecolor{gray33}{rgb}{0.33,0.33,0.33}
\definecolor{gray34}{rgb}{0.34,0.34,0.34}
\definecolor{gray35}{rgb}{0.35,0.35,0.35}
\definecolor{gray36}{rgb}{0.36,0.36,0.36}
\definecolor{gray37}{rgb}{0.37,0.37,0.37}
\definecolor{gray38}{rgb}{0.38,0.38,0.38}
\definecolor{gray39}{rgb}{0.39,0.39,0.39}
\definecolor{gray3}{rgb}{0.03,0.03,0.03}
\definecolor{gray40}{rgb}{0.40,0.40,0.40}
\definecolor{gray41}{rgb}{0.41,0.41,0.41}
\definecolor{gray42}{rgb}{0.42,0.42,0.42}
\definecolor{gray43}{rgb}{0.43,0.43,0.43}
\definecolor{gray44}{rgb}{0.44,0.44,0.44}
\definecolor{gray45}{rgb}{0.45,0.45,0.45}
\definecolor{gray46}{rgb}{0.46,0.46,0.46}
\definecolor{gray47}{rgb}{0.47,0.47,0.47}
\definecolor{gray48}{rgb}{0.48,0.48,0.48}
\definecolor{gray49}{rgb}{0.49,0.49,0.49}
\definecolor{gray4}{rgb}{0.04,0.04,0.04}
\definecolor{gray50}{rgb}{0.50,0.50,0.50}
\definecolor{gray51}{rgb}{0.51,0.51,0.51}
\definecolor{gray52}{rgb}{0.52,0.52,0.52}
\definecolor{gray53}{rgb}{0.53,0.53,0.53}
\definecolor{gray54}{rgb}{0.54,0.54,0.54}
\definecolor{gray55}{rgb}{0.55,0.55,0.55}
\definecolor{gray56}{rgb}{0.56,0.56,0.56}
\definecolor{gray57}{rgb}{0.57,0.57,0.57}
\definecolor{gray58}{rgb}{0.58,0.58,0.58}
\definecolor{gray59}{rgb}{0.59,0.59,0.59}
\definecolor{gray5}{rgb}{0.05,0.05,0.05}
\definecolor{gray60}{rgb}{0.60,0.60,0.60}
\definecolor{gray61}{rgb}{0.61,0.61,0.61}
\definecolor{gray62}{rgb}{0.62,0.62,0.62}
\definecolor{gray63}{rgb}{0.63,0.63,0.63}
\definecolor{gray64}{rgb}{0.64,0.64,0.64}
\definecolor{gray65}{rgb}{0.65,0.65,0.65}
\definecolor{gray66}{rgb}{0.66,0.66,0.66}
\definecolor{gray67}{rgb}{0.67,0.67,0.67}
\definecolor{gray68}{rgb}{0.68,0.68,0.68}
\definecolor{gray69}{rgb}{0.69,0.69,0.69}
\definecolor{gray6}{rgb}{0.06,0.06,0.06}
\definecolor{gray70}{rgb}{0.70,0.70,0.70}
\definecolor{gray71}{rgb}{0.71,0.71,0.71}
\definecolor{gray72}{rgb}{0.72,0.72,0.72}
\definecolor{gray73}{rgb}{0.73,0.73,0.73}
\definecolor{gray74}{rgb}{0.74,0.74,0.74}
\definecolor{gray75}{rgb}{0.75,0.75,0.75}
\definecolor{gray76}{rgb}{0.76,0.76,0.76}
\definecolor{gray77}{rgb}{0.77,0.77,0.77}
\definecolor{gray78}{rgb}{0.78,0.78,0.78}
\definecolor{gray79}{rgb}{0.79,0.79,0.79}
\definecolor{gray7}{rgb}{0.07,0.07,0.07}
\definecolor{gray80}{rgb}{0.80,0.80,0.80}
\definecolor{gray81}{rgb}{0.81,0.81,0.81}
\definecolor{gray82}{rgb}{0.82,0.82,0.82}
\definecolor{gray83}{rgb}{0.83,0.83,0.83}
\definecolor{gray84}{rgb}{0.84,0.84,0.84}
\definecolor{gray85}{rgb}{0.85,0.85,0.85}
\definecolor{gray86}{rgb}{0.86,0.86,0.86}
\definecolor{gray87}{rgb}{0.87,0.87,0.87}
\definecolor{gray88}{rgb}{0.88,0.88,0.88}
\definecolor{gray89}{rgb}{0.89,0.89,0.89}
\definecolor{gray8}{rgb}{0.08,0.08,0.08}
\definecolor{gray90}{rgb}{0.90,0.90,0.90}
\definecolor{gray91}{rgb}{0.91,0.91,0.91}
\definecolor{gray92}{rgb}{0.92,0.92,0.92}
\definecolor{gray93}{rgb}{0.93,0.93,0.93}
\definecolor{gray94}{rgb}{0.94,0.94,0.94}
\definecolor{gray95}{rgb}{0.95,0.95,0.95}
\definecolor{gray96}{rgb}{0.96,0.96,0.96}
\definecolor{gray97}{rgb}{0.97,0.97,0.97}
\definecolor{gray98}{rgb}{0.98,0.98,0.98}
\definecolor{gray99}{rgb}{0.99,0.99,0.99}
\definecolor{gray9}{rgb}{0.09,0.09,0.09}
\definecolor{gray}{rgb}{0.75,0.75,0.75}
\definecolor{green1}{rgb}{0.00,1.00,0.00}
\definecolor{green2}{rgb}{0.00,0.93,0.00}
\definecolor{green3}{rgb}{0.00,0.80,0.00}
\definecolor{green4}{rgb}{0.00,0.55,0.00}
\definecolor{greenyellow}{rgb}{0.68,1.00,0.18}
\definecolor{green}{rgb}{0.00,1.00,0.00}
\definecolor{grey0}{rgb}{0.00,0.00,0.00}
\definecolor{grey100}{rgb}{1.00,1.00,1.00}
\definecolor{grey10}{rgb}{0.10,0.10,0.10}
\definecolor{grey11}{rgb}{0.11,0.11,0.11}
\definecolor{grey12}{rgb}{0.12,0.12,0.12}
\definecolor{grey13}{rgb}{0.13,0.13,0.13}
\definecolor{grey14}{rgb}{0.14,0.14,0.14}
\definecolor{grey15}{rgb}{0.15,0.15,0.15}
\definecolor{grey16}{rgb}{0.16,0.16,0.16}
\definecolor{grey17}{rgb}{0.17,0.17,0.17}
\definecolor{grey18}{rgb}{0.18,0.18,0.18}
\definecolor{grey19}{rgb}{0.19,0.19,0.19}
\definecolor{grey1}{rgb}{0.01,0.01,0.01}
\definecolor{grey20}{rgb}{0.20,0.20,0.20}
\definecolor{grey21}{rgb}{0.21,0.21,0.21}
\definecolor{grey22}{rgb}{0.22,0.22,0.22}
\definecolor{grey23}{rgb}{0.23,0.23,0.23}
\definecolor{grey24}{rgb}{0.24,0.24,0.24}
\definecolor{grey25}{rgb}{0.25,0.25,0.25}
\definecolor{grey26}{rgb}{0.26,0.26,0.26}
\definecolor{grey27}{rgb}{0.27,0.27,0.27}
\definecolor{grey28}{rgb}{0.28,0.28,0.28}
\definecolor{grey29}{rgb}{0.29,0.29,0.29}
\definecolor{grey2}{rgb}{0.02,0.02,0.02}
\definecolor{grey30}{rgb}{0.30,0.30,0.30}
\definecolor{grey31}{rgb}{0.31,0.31,0.31}
\definecolor{grey32}{rgb}{0.32,0.32,0.32}
\definecolor{grey33}{rgb}{0.33,0.33,0.33}
\definecolor{grey34}{rgb}{0.34,0.34,0.34}
\definecolor{grey35}{rgb}{0.35,0.35,0.35}
\definecolor{grey36}{rgb}{0.36,0.36,0.36}
\definecolor{grey37}{rgb}{0.37,0.37,0.37}
\definecolor{grey38}{rgb}{0.38,0.38,0.38}
\definecolor{grey39}{rgb}{0.39,0.39,0.39}
\definecolor{grey3}{rgb}{0.03,0.03,0.03}
\definecolor{grey40}{rgb}{0.40,0.40,0.40}
\definecolor{grey41}{rgb}{0.41,0.41,0.41}
\definecolor{grey42}{rgb}{0.42,0.42,0.42}
\definecolor{grey43}{rgb}{0.43,0.43,0.43}
\definecolor{grey44}{rgb}{0.44,0.44,0.44}
\definecolor{grey45}{rgb}{0.45,0.45,0.45}
\definecolor{grey46}{rgb}{0.46,0.46,0.46}
\definecolor{grey47}{rgb}{0.47,0.47,0.47}
\definecolor{grey48}{rgb}{0.48,0.48,0.48}
\definecolor{grey49}{rgb}{0.49,0.49,0.49}
\definecolor{grey4}{rgb}{0.04,0.04,0.04}
\definecolor{grey50}{rgb}{0.50,0.50,0.50}
\definecolor{grey51}{rgb}{0.51,0.51,0.51}
\definecolor{grey52}{rgb}{0.52,0.52,0.52}
\definecolor{grey53}{rgb}{0.53,0.53,0.53}
\definecolor{grey54}{rgb}{0.54,0.54,0.54}
\definecolor{grey55}{rgb}{0.55,0.55,0.55}
\definecolor{grey56}{rgb}{0.56,0.56,0.56}
\definecolor{grey57}{rgb}{0.57,0.57,0.57}
\definecolor{grey58}{rgb}{0.58,0.58,0.58}
\definecolor{grey59}{rgb}{0.59,0.59,0.59}
\definecolor{grey5}{rgb}{0.05,0.05,0.05}
\definecolor{grey60}{rgb}{0.60,0.60,0.60}
\definecolor{grey61}{rgb}{0.61,0.61,0.61}
\definecolor{grey62}{rgb}{0.62,0.62,0.62}
\definecolor{grey63}{rgb}{0.63,0.63,0.63}
\definecolor{grey64}{rgb}{0.64,0.64,0.64}
\definecolor{grey65}{rgb}{0.65,0.65,0.65}
\definecolor{grey66}{rgb}{0.66,0.66,0.66}
\definecolor{grey67}{rgb}{0.67,0.67,0.67}
\definecolor{grey68}{rgb}{0.68,0.68,0.68}
\definecolor{grey69}{rgb}{0.69,0.69,0.69}
\definecolor{grey6}{rgb}{0.06,0.06,0.06}
\definecolor{grey70}{rgb}{0.70,0.70,0.70}
\definecolor{grey71}{rgb}{0.71,0.71,0.71}
\definecolor{grey72}{rgb}{0.72,0.72,0.72}
\definecolor{grey73}{rgb}{0.73,0.73,0.73}
\definecolor{grey74}{rgb}{0.74,0.74,0.74}
\definecolor{grey75}{rgb}{0.75,0.75,0.75}
\definecolor{grey76}{rgb}{0.76,0.76,0.76}
\definecolor{grey77}{rgb}{0.77,0.77,0.77}
\definecolor{grey78}{rgb}{0.78,0.78,0.78}
\definecolor{grey79}{rgb}{0.79,0.79,0.79}
\definecolor{grey7}{rgb}{0.07,0.07,0.07}
\definecolor{grey80}{rgb}{0.80,0.80,0.80}
\definecolor{grey81}{rgb}{0.81,0.81,0.81}
\definecolor{grey82}{rgb}{0.82,0.82,0.82}
\definecolor{grey83}{rgb}{0.83,0.83,0.83}
\definecolor{grey84}{rgb}{0.84,0.84,0.84}
\definecolor{grey85}{rgb}{0.85,0.85,0.85}
\definecolor{grey86}{rgb}{0.86,0.86,0.86}
\definecolor{grey87}{rgb}{0.87,0.87,0.87}
\definecolor{grey88}{rgb}{0.88,0.88,0.88}
\definecolor{grey89}{rgb}{0.89,0.89,0.89}
\definecolor{grey8}{rgb}{0.08,0.08,0.08}
\definecolor{grey90}{rgb}{0.90,0.90,0.90}
\definecolor{grey91}{rgb}{0.91,0.91,0.91}
\definecolor{grey92}{rgb}{0.92,0.92,0.92}
\definecolor{grey93}{rgb}{0.93,0.93,0.93}
\definecolor{grey94}{rgb}{0.94,0.94,0.94}
\definecolor{grey95}{rgb}{0.95,0.95,0.95}
\definecolor{grey96}{rgb}{0.96,0.96,0.96}
\definecolor{grey97}{rgb}{0.97,0.97,0.97}
\definecolor{grey98}{rgb}{0.98,0.98,0.98}
\definecolor{grey99}{rgb}{0.99,0.99,0.99}
\definecolor{grey9}{rgb}{0.09,0.09,0.09}
\definecolor{grey}{rgb}{0.75,0.75,0.75}
\definecolor{honeydew1}{rgb}{0.94,1.00,0.94}
\definecolor{honeydew2}{rgb}{0.88,0.93,0.88}
\definecolor{honeydew3}{rgb}{0.76,0.80,0.76}
\definecolor{honeydew4}{rgb}{0.51,0.55,0.51}
\definecolor{honeydew}{rgb}{0.94,1.00,0.94}
\definecolor{hotpink}{rgb}{1.00,0.41,0.71}
\definecolor{indianred}{rgb}{0.80,0.36,0.36}
\definecolor{ivory1}{rgb}{1.00,1.00,0.94}
\definecolor{ivory2}{rgb}{0.93,0.93,0.88}
\definecolor{ivory3}{rgb}{0.80,0.80,0.76}
\definecolor{ivory4}{rgb}{0.55,0.55,0.51}
\definecolor{ivory}{rgb}{1.00,1.00,0.94}
\definecolor{khaki1}{rgb}{1.00,0.96,0.56}
\definecolor{khaki2}{rgb}{0.93,0.90,0.52}
\definecolor{khaki3}{rgb}{0.80,0.78,0.45}
\definecolor{khaki4}{rgb}{0.55,0.53,0.31}
\definecolor{khaki}{rgb}{0.94,0.90,0.55}
\definecolor{lavenderblush}{rgb}{1.00,0.94,0.96}
\definecolor{lavender}{rgb}{0.90,0.90,0.98}
\definecolor{lawngreen}{rgb}{0.49,0.99,0.00}
\definecolor{lemonchiffon}{rgb}{1.00,0.98,0.80}
\definecolor{lightblue}{rgb}{0.68,0.85,0.90}
\definecolor{lightcoral}{rgb}{0.94,0.50,0.50}
\definecolor{lightcyan}{rgb}{0.88,1.00,1.00}
\definecolor{lightgoldenrod}{rgb}{0.93,0.87,0.51}
\definecolor{lightgoldenrod}{rgb}{0.98,0.98,0.82}
\definecolor{lightgray}{rgb}{0.83,0.83,0.83}
\definecolor{lightgreen}{rgb}{0.56,0.93,0.56}
\definecolor{lightgrey}{rgb}{0.83,0.83,0.83}
\definecolor{lightpink}{rgb}{1.00,0.71,0.76}
\definecolor{lightsalmon}{rgb}{1.00,0.63,0.48}
\definecolor{lightsea}{rgb}{0.13,0.70,0.67}
\definecolor{lightsky}{rgb}{0.53,0.81,0.98}
\definecolor{lightslate}{rgb}{0.47,0.53,0.60}
\definecolor{lightslate}{rgb}{0.47,0.53,0.60}
\definecolor{lightslate}{rgb}{0.52,0.44,1.00}
\definecolor{lightsteel}{rgb}{0.69,0.77,0.87}
\definecolor{lightyellow}{rgb}{1.00,1.00,0.88}
\definecolor{limegreen}{rgb}{0.20,0.80,0.20}
\definecolor{linen}{rgb}{0.98,0.94,0.90}
\definecolor{magenta1}{rgb}{1.00,0.00,1.00}
\definecolor{magenta2}{rgb}{0.93,0.00,0.93}
\definecolor{magenta3}{rgb}{0.80,0.00,0.80}
\definecolor{magenta4}{rgb}{0.55,0.00,0.55}
\definecolor{magenta}{rgb}{1.00,0.00,1.00}
\definecolor{maroon1}{rgb}{1.00,0.20,0.70}
\definecolor{maroon2}{rgb}{0.93,0.19,0.65}
\definecolor{maroon3}{rgb}{0.80,0.16,0.56}
\definecolor{maroon4}{rgb}{0.55,0.11,0.38}
\definecolor{maroon}{rgb}{0.69,0.19,0.38}
\definecolor{mediumaquamarine}{rgb}{0.40,0.80,0.67}
\definecolor{mediumblue}{rgb}{0.00,0.00,0.80}
\definecolor{mediumorchid}{rgb}{0.73,0.33,0.83}
\definecolor{mediumpurple}{rgb}{0.58,0.44,0.86}
\definecolor{mediumsea}{rgb}{0.24,0.70,0.44}
\definecolor{mediumslate}{rgb}{0.48,0.41,0.93}
\definecolor{mediumspring}{rgb}{0.00,0.98,0.60}
\definecolor{mediumturquoise}{rgb}{0.28,0.82,0.80}
\definecolor{mediumviolet}{rgb}{0.78,0.08,0.52}
\definecolor{midnightblue}{rgb}{0.10,0.10,0.44}
\definecolor{mintcream}{rgb}{0.96,1.00,0.98}
\definecolor{mistyrose}{rgb}{1.00,0.89,0.88}
\definecolor{moccasin}{rgb}{1.00,0.89,0.71}
\definecolor{navajowhite}{rgb}{1.00,0.87,0.68}
\definecolor{navyblue}{rgb}{0.00,0.00,0.50}
\definecolor{navy}{rgb}{0.00,0.00,0.50}
\definecolor{oldlace}{rgb}{0.99,0.96,0.90}
\definecolor{olivedrab}{rgb}{0.42,0.56,0.14}
\definecolor{orange1}{rgb}{1.00,0.65,0.00}
\definecolor{orange2}{rgb}{0.93,0.60,0.00}
\definecolor{orange3}{rgb}{0.80,0.52,0.00}
\definecolor{orange4}{rgb}{0.55,0.35,0.00}
\definecolor{orangered}{rgb}{1.00,0.27,0.00}
\definecolor{orange}{rgb}{1.00,0.65,0.00}
\definecolor{orchid1}{rgb}{1.00,0.51,0.98}
\definecolor{orchid2}{rgb}{0.93,0.48,0.91}
\definecolor{orchid3}{rgb}{0.80,0.41,0.79}
\definecolor{orchid4}{rgb}{0.55,0.28,0.54}
\definecolor{orchid}{rgb}{0.85,0.44,0.84}
\definecolor{palegoldenrod}{rgb}{0.93,0.91,0.67}
\definecolor{palegreen}{rgb}{0.60,0.98,0.60}
\definecolor{paleturquoise}{rgb}{0.69,0.93,0.93}
\definecolor{paleviolet}{rgb}{0.86,0.44,0.58}
\definecolor{papayawhip}{rgb}{1.00,0.94,0.84}
\definecolor{peachpuff}{rgb}{1.00,0.85,0.73}
\definecolor{peru}{rgb}{0.80,0.52,0.25}
\definecolor{pink1}{rgb}{1.00,0.71,0.77}
\definecolor{pink2}{rgb}{0.93,0.66,0.72}
\definecolor{pink3}{rgb}{0.80,0.57,0.62}
\definecolor{pink4}{rgb}{0.55,0.39,0.42}
\definecolor{pink}{rgb}{1.00,0.75,0.80}
\definecolor{plum1}{rgb}{1.00,0.73,1.00}
\definecolor{plum2}{rgb}{0.93,0.68,0.93}
\definecolor{plum3}{rgb}{0.80,0.59,0.80}
\definecolor{plum4}{rgb}{0.55,0.40,0.55}
\definecolor{plum}{rgb}{0.87,0.63,0.87}
\definecolor{powderblue}{rgb}{0.69,0.88,0.90}
\definecolor{purple1}{rgb}{0.61,0.19,1.00}
\definecolor{purple2}{rgb}{0.57,0.17,0.93}
\definecolor{purple3}{rgb}{0.49,0.15,0.80}
\definecolor{purple4}{rgb}{0.33,0.10,0.55}
\definecolor{purple}{rgb}{0.63,0.13,0.94}
\definecolor{red1}{rgb}{1.00,0.00,0.00}
\definecolor{red2}{rgb}{0.93,0.00,0.00}
\definecolor{red3}{rgb}{0.80,0.00,0.00}
\definecolor{red4}{rgb}{0.55,0.00,0.00}
\definecolor{red}{rgb}{1.00,0.00,0.00}
\definecolor{rosybrown}{rgb}{0.74,0.56,0.56}
\definecolor{royalblue}{rgb}{0.25,0.41,0.88}
\definecolor{saddlebrown}{rgb}{0.55,0.27,0.07}
\definecolor{salmon1}{rgb}{1.00,0.55,0.41}
\definecolor{salmon2}{rgb}{0.93,0.51,0.38}
\definecolor{salmon3}{rgb}{0.80,0.44,0.33}
\definecolor{salmon4}{rgb}{0.55,0.30,0.22}
\definecolor{salmon}{rgb}{0.98,0.50,0.45}
\definecolor{sandybrown}{rgb}{0.96,0.64,0.38}
\definecolor{seagreen}{rgb}{0.18,0.55,0.34}
\definecolor{seashell1}{rgb}{1.00,0.96,0.93}
\definecolor{seashell2}{rgb}{0.93,0.90,0.87}
\definecolor{seashell3}{rgb}{0.80,0.77,0.75}
\definecolor{seashell4}{rgb}{0.55,0.53,0.51}
\definecolor{seashell}{rgb}{1.00,0.96,0.93}
\definecolor{sienna1}{rgb}{1.00,0.51,0.28}
\definecolor{sienna2}{rgb}{0.93,0.47,0.26}
\definecolor{sienna3}{rgb}{0.80,0.41,0.22}
\definecolor{sienna4}{rgb}{0.55,0.28,0.15}
\definecolor{sienna}{rgb}{0.63,0.32,0.18}
\definecolor{skyblue}{rgb}{0.53,0.81,0.92}
\definecolor{slateblue}{rgb}{0.42,0.35,0.80}
\definecolor{slategray}{rgb}{0.44,0.50,0.56}
\definecolor{slategrey}{rgb}{0.44,0.50,0.56}
\definecolor{snow1}{rgb}{1.00,0.98,0.98}
\definecolor{snow2}{rgb}{0.93,0.91,0.91}
\definecolor{snow3}{rgb}{0.80,0.79,0.79}
\definecolor{snow4}{rgb}{0.55,0.54,0.54}
\definecolor{snow}{rgb}{1.00,0.98,0.98}
\definecolor{springgreen}{rgb}{0.00,1.00,0.50}
\definecolor{steelblue}{rgb}{0.27,0.51,0.71}
\definecolor{tan1}{rgb}{1.00,0.65,0.31}
\definecolor{tan2}{rgb}{0.93,0.60,0.29}
\definecolor{tan3}{rgb}{0.80,0.52,0.25}
\definecolor{tan4}{rgb}{0.55,0.35,0.17}
\definecolor{tan}{rgb}{0.82,0.71,0.55}
\definecolor{thistle1}{rgb}{1.00,0.88,1.00}
\definecolor{thistle2}{rgb}{0.93,0.82,0.93}
\definecolor{thistle3}{rgb}{0.80,0.71,0.80}
\definecolor{thistle4}{rgb}{0.55,0.48,0.55}
\definecolor{thistle}{rgb}{0.85,0.75,0.85}
\definecolor{tomato1}{rgb}{1.00,0.39,0.28}
\definecolor{tomato2}{rgb}{0.93,0.36,0.26}
\definecolor{tomato3}{rgb}{0.80,0.31,0.22}
\definecolor{tomato4}{rgb}{0.55,0.21,0.15}
\definecolor{tomato}{rgb}{1.00,0.39,0.28}
\definecolor{turquoise1}{rgb}{0.00,0.96,1.00}
\definecolor{turquoise2}{rgb}{0.00,0.90,0.93}
\definecolor{turquoise3}{rgb}{0.00,0.77,0.80}
\definecolor{turquoise4}{rgb}{0.00,0.53,0.55}
\definecolor{turquoise}{rgb}{0.25,0.88,0.82}
\definecolor{violetred}{rgb}{0.82,0.13,0.56}
\definecolor{violet}{rgb}{0.93,0.51,0.93}
\definecolor{wheat1}{rgb}{1.00,0.91,0.73}
\definecolor{wheat2}{rgb}{0.93,0.85,0.68}
\definecolor{wheat3}{rgb}{0.80,0.73,0.59}
\definecolor{wheat4}{rgb}{0.55,0.49,0.40}
\definecolor{wheat}{rgb}{0.96,0.87,0.70}
\definecolor{whitesmoke}{rgb}{0.96,0.96,0.96}
\definecolor{white}{rgb}{1.00,1.00,1.00}
\definecolor{yellow1}{rgb}{1.00,1.00,0.00}
\definecolor{yellow2}{rgb}{0.93,0.93,0.00}
\definecolor{yellow3}{rgb}{0.80,0.80,0.00}
\definecolor{yellow4}{rgb}{0.55,0.55,0.00}
\definecolor{yellowgreen}{rgb}{0.60,0.80,0.20}
\definecolor{yellow}{rgb}{1.00,1.00,0.00}